# Extraordinary momentum and spin in evanescent waves

Konstantin Y. Bliokh[1,2], Aleksandr Y. Bekshaev[3,4], and Franco Nori[3,5,6]

[1]*iTHES Research Group, RIKEN, Wako-shi, Saitama 351-0198, Japan*
[2]*A. Usikov Institute of Radiophysics and Electronics, NASU, Kharkov 61085, Ukraine*
[3]*Center for Emergent Matter Science, RIKEN, Wako-shi, Saitama 351-0198, Japan*
[4]*I. I. Mechnikov National University, Dvorianska 2, Odessa, 65082, Ukraine*
[5]*Physics Department, University of Michigan, Ann Arbor, Michigan 48109-1040, USA*
[6]*Department of Physics, Korea University, Seoul 136-713, Korea*

Momentum and spin represent fundamental dynamical properties of quantum particles and fields. In particular, propagating optical waves (photons) carry momentum and longitudinal spin determined by the wave vector and circular polarization, respectively. Here we show that exactly the opposite can be the case for evanescent optical waves. A single evanescent wave possesses a spin component, which is independent of the polarization and is orthogonal to the wave vector. Furthermore, such a wave carries a momentum component, which is determined by the circular polarization and is also orthogonal to the wave vector. We show that these extraordinary properties reveal a fundamental Belinfante's spin momentum, known in field theory and unobservable in propagating fields. We demonstrate that the transverse momentum and spin push and twist a probe Mie particle in an evanescent field. This allows the observation of 'impossible' properties of light and of a fundamental field-theory quantity, which was previously considered as 'virtual'.

**Introduction**

It has been known for more than a century, since the seminal works by J.H. Poynting [1], that light carries momentum and angular momentum (AM) [2,3]. These are the main dynamical properties of electromagnetic waves, which are also preserved in the quantum-mechanical picture of photons [4]. Optical momentum and AM play a crucial role in various light-matter interactions [5–7], including laser cooling [8–10], optical manipulation of atoms or small particles [8–13], and in optomechanical systems [14].

The simplest example of an optical field carrying momentum and spin AM is an elliptically-polarized plane wave. Assuming free-space propagation along the $z$-axis, the complex electric field of such wave can be written as

$$\mathbf{E}(\mathbf{r}) = A \frac{(\hat{\mathbf{x}} + m\hat{\mathbf{y}})}{\sqrt{1+|m|^2}} \exp(ikz). \qquad (1)$$

Here $A$ is the wave amplitude, $\hat{\mathbf{x}}$ and $\hat{\mathbf{y}}$ are unit vectors of the corresponding axes, the complex number $m$ determines the polarization state with $\sigma = \dfrac{2\operatorname{Im} m}{1+|m|^2} \in [-1,1]$ being the helicity (ellipticity of polarization), $k = \omega/c$ is the wave number, and throughout the paper we imply monochromatic fields with omitted $\exp(-i\omega t)$ factor.

The momentum and spin AM in wave (1) can be characterized by the corresponding spatial densities

$$\mathbf{p} = \frac{w}{\omega} k\,\hat{\mathbf{z}}, \quad \mathbf{s} = \frac{w}{\omega} \sigma\,\hat{\mathbf{z}}, \qquad (2)$$

where $w = \gamma \omega |A|^2$ is the energy density and we use Gaussian units with $\gamma = (8\pi\omega)^{-1}$. In agreement with the quantum-mechanical picture of photons [4], the momentum $\mathbf{p}$ is determined by the wave vector $k\hat{\mathbf{z}}$ and is independent of polarization. At the same time, the spin $\mathbf{s}$ is proportional to the polarization helicity $\sigma$ and is also collinear with the wave vector.

The optical momentum and spin densities can be measured experimentally by placing a small absorbing particle in the field and observing its linear and spinning motion [15–19]. Naturally, the radiation force and torque (with respect to the particle's centre) quantify the momentum and AM transfer to the particle and are proportional to the densities (2) in the plane wave (1): $\mathbf{F} \propto \mathbf{p}$ and $\mathbf{T} \propto \mathbf{s}$ [19–25], Fig. 1.

The above picture is simple and intuitively clear in the plane-wave case, but complex, spatially inhomogeneous fields require a more careful approach. Below we show that one of the simplest examples of an inhomogeneous field – a single evanescent wave – exhibits extraordinary momentum and spin properties, which are in sharp contrast to what is known about photons, optical momentum, and spin.

## Results

**Momentum and spin densities from field theory.** The momentum density $\mathbf{p}(\mathbf{r})$ of a quantum or classical wave field appears in the energy-momentum tensor within the corresponding field theory [26], where momentum density also represents the energy flux density. For scalar fields, the momentum density can be written as a local expectation value of the canonical momentum operator $\hat{\mathbf{p}} = -i\nabla$, i.e., $\mathbf{p} = \mathrm{Re}(\psi^\dagger \hat{\mathbf{p}} \psi)$, where $\psi(\mathbf{r})$ is the wave function, and we use units $\hbar = 1$. However, for vector fields, an additional spin momentum density was introduced in 1939 by F.J. Belinfante [27] to explain the spin of quantum particles and symmetrize the canonical energy-momentum tensor in field theory. The spin momentum is added to the canonical (or orbital) momentum density, resulting in [26–31]:

$$\mathbf{p} = \mathrm{Re}(\vec{\psi}^\dagger \hat{\mathbf{p}} \vec{\psi}) + \frac{1}{2}\nabla \times \mathbf{s} \equiv \mathbf{p}^O + \mathbf{p}^S. \qquad (3)$$

Here $\vec{\psi}(\mathbf{r})$ is the spinor wave function, whereas $\mathbf{s}(\mathbf{r})$ is the spin AM density defined as the local expectation value of the corresponding matrix spin operator $\hat{\mathbf{S}}$:

$$\mathbf{s} = \vec{\psi}^\dagger \hat{\mathbf{S}} \vec{\psi}. \qquad (4)$$

Equations (3) and (4) are fundamental and hold true for various particles. For Dirac electron, $\hat{\mathbf{S}}$ is the spin-1/2 operator and $\vec{\psi}(\mathbf{r},t)$ is the Dirac bi-spinor [28–31]. In the case of monochromatic electromagnetic waves (photons), equation (3) yields the time-averaged Poynting vector $\mathbf{p} = \gamma k\, \mathrm{Re}(\mathbf{E}^* \times \mathbf{H})$, when $\hat{\mathbf{S}}$ is given by spin-1 matrices which act on the complex electric and magnetic field amplitudes, $\vec{\psi}(\mathbf{r}) = \sqrt{\dfrac{\gamma}{2}} \begin{pmatrix} \mathbf{E}(\mathbf{r}) \\ \mathbf{H}(\mathbf{r}) \end{pmatrix}$, supplied with the free-space Maxwell equations [22,26–31] (see Supplementary Note 1). Explicitly, the optical momentum and spin densities (3) and (4) read

$$\mathbf{p}^O = \frac{\gamma}{2}\mathrm{Im}\left[\mathbf{E}^* \cdot (\nabla)\mathbf{E} + \mathbf{H}^* \cdot (\nabla)\mathbf{H}\right], \quad \mathbf{p}^S = \frac{1}{2}\nabla \times \mathbf{s},$$

$$\mathbf{s} = \frac{\gamma}{2}\mathrm{Im}\left[\mathbf{E}^* \times \mathbf{E} + \mathbf{H}^* \times \mathbf{H}\right]. \qquad (5)$$

Note that these quantities are naturally split into electric- and magnetic-field contributions: $\mathbf{p}^O = \mathbf{p}_e^O + \mathbf{p}_m^O$ and $\mathbf{s} = \mathbf{s}_e + \mathbf{s}_m$. It is important to emphasize that although the Poynting vector $\mathbf{p}$ is usually considered in optics as a single momentum density of light [2], it actually represents the sum of two quantities $\mathbf{p}^O$ and $\mathbf{p}^S$, with drastically different physical meanings and properties. Below we uncover the contrasting manifestations of the canonical and spin momenta.

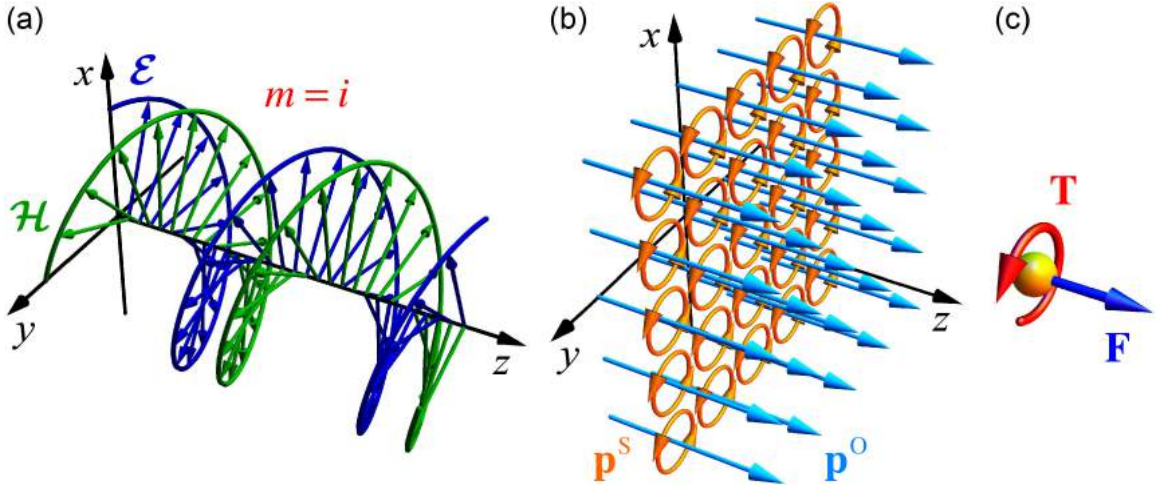

**Figure 1. Momentum and spin in a circularly-polarized propagating plane wave.** The complex wave electric field is given by equation (1) with $m = i$, i.e. $\sigma = 1$. **(a)** Instantaneous electric and magnetic fields, $\mathcal{E}(\mathbf{r},t) = \text{Re}\left[\mathbf{E}(\mathbf{r})e^{-i\omega t}\right]$ and $\mathcal{H}(\mathbf{r},t) = \text{Re}\left[\mathbf{H}(\mathbf{r})e^{-i\omega t}\right]$, form helical distributions (see Supplementary Note 2 and Supplementary Figure 1). As the wave propagates along the $z$-axis, the fields rotate in the transverse $(x,y)$ plane. This rotation generates the spin AM density $\mathbf{s} \propto \sigma \hat{\mathbf{z}}$, which is represented in **(b)** by multiple loops of the (zero-net) spin-momentum $\mathbf{p}^S = \nabla \times \mathbf{s}/2$ in the transverse plane. At the same time, the wave propagation produces the canonical (orbital) momentum density $\mathbf{p}^O \propto k\hat{\mathbf{z}}$. **(c)** Orbital momentum and spin AM are locally transferred to a probe particle, thereby exerting a radiation force $\mathbf{F} \propto \mathbf{p}^O$ and torque $\mathbf{T} \propto \mathbf{s}$ on it, Eqs. (6).

The canonical (orbital) and spin parts of the momentum density (3), $\mathbf{p} = \mathbf{p}^O + \mathbf{p}^S$, generate, respectively, the orbital and spin parts of the AM density: $\mathbf{j} = \mathbf{l} + \mathbf{s}$ [26,27,32,33]. The orbital AM density is $\mathbf{l} = \mathbf{r} \times \mathbf{p}^O$, and this is an extrinsic origin-dependent quantity. At the same time, the spin AM density $\mathbf{s}$, Eq. (4), is intrinsic (origin-independent). Nonetheless, its integral value is determined by the circulation of the spin momentum: $\mathbf{S} = \int \mathbf{s}\, dV = \int \mathbf{r} \times \mathbf{p}^S dV$, where integration by parts should be performed [28–33].

The orbital momentum density $\mathbf{p}^O$ is naturally proportional to the local phase gradient (wave vector) in the field [22]. In contrast, Belinfante's spin momentum $\mathbf{p}^S$ is a rather enigmatic quantity [26–33]. On the one hand, the spin momentum provides the physical origin of the spin AM of quantum particles. On the other hand, it is usually considered as an auxiliary 'virtual' quantity, which cannot be observed *per se*. Indeed, the spin momentum represents a solenoidal current, which does not contribute to the energy transport ($\nabla \cdot \mathbf{p}^S = 0$ and $\int \mathbf{p}^S dV = 0$), and only generates spin AM. Consider, for instance, the elliptically-polarized electromagnetic plane wave (1). This field carries only the longitudinal orbital momentum density: $\mathbf{p}^O \propto k\hat{\mathbf{z}}$, while the spin momentum vanishes: $\mathbf{p}^S = 0$. In this case, what generates the spin AM density $\mathbf{s} \propto \sigma \hat{\mathbf{z}}$? This known paradox [34,35] is resolved by representing the zero transverse momentum as an array of infinitely small loops of circulating spin momentum in the $(x,y)$ plane [26,30–32], see Fig. 1b. Currents from the neighbouring loops cancel each other, but at the same time they provide non-zero circulation along any finite closed loop, i.e., non-zero spin AM along the $z$-axis. The formal integral relation between $\mathbf{p}^S$ and $\mathbf{s}$ does not work here, because a plane wave is an unbounded state; the introduction of a boundary (e.g., Gaussian intensity distribution in the transverse plane) immediately produces a non-zero boundary spin current $\mathbf{p}^S \neq 0$ with the integral circulation yielding the spin $\mathbf{S}$ [28,30–32]. Thus, Belinfante's spin momentum is similar to the boundary magnetization current or topological quantum-Hall current in solid-state systems (multiple current loops are produced there by electron orbitals), whereas the spin AM is analogous to the bulk magnetization in such systems.

**Measurements of the momentum and spin densities via probe particles.** Having the above theoretical picture, let us consider measurements of the momentum and spin densities in an electromagnetic field. As we mentioned in the Introduction, a small absorbing particle immersed in the field can be employed as a natural meter of these quantities. Calculating the radiation force and torque on a dipole Rayleigh particle with equal electric and magnetic polarizabilities, $\alpha_e = \alpha_m = \alpha$, one can show that they are determined by the canonical momentum and spin densities [19–25] (see Supplementary Note 3):

$$\mathbf{F} = \gamma^{-1}\left[\frac{1}{2\omega}\text{Re}(\alpha)\nabla w + \text{Im}(\alpha)\mathbf{p}^O\right], \quad \mathbf{T} = \gamma^{-1}\text{Im}(\alpha)\mathbf{s}. \qquad (6)$$

Here $w = \frac{\gamma}{2}\omega\left(|\mathbf{E}|^2 + |\mathbf{H}|^2\right) = w_e + w_m$ is the energy density of the field. The first term in the first equation (6) is the gradient force, while the second term is the radiation-pressure force. Together they 'measure' the imaginary and real parts of the complex canonical momentum: $\tilde{\mathbf{p}}^O(\mathbf{r}) = \vec{\psi}^\dagger(\mathbf{r})\hat{\mathbf{p}}\vec{\psi}(\mathbf{r}) = \mathbf{p}^O(\mathbf{r}) - i\nabla w(\mathbf{r})/2\omega$, which is proportional to the quantum weak value of the photon's momentum, $\mathbf{p}_w = \langle\mathbf{r}|\hat{\mathbf{p}}|\vec{\psi}\rangle/\langle\mathbf{r}|\vec{\psi}\rangle$ [22,19,36].

Importantly, it is the canonical momentum $\mathbf{p}^O$ rather than the Poynting vector $\mathbf{p}$ that represents the physically-meaningful momentum density of light that appears in experiments ($\mathbf{p} = \mathbf{p}^O$ only in linearly-polarized paraxial fields and plane waves) [19,33]. In particular, the orbital and spinning motions of probe particles in circularly-polarized vortex beams [15–18] originate exactly due to the force from the azimuthal component of $\mathbf{p}^O$ and the torque from the longitudinal component of the spin density $\mathbf{s}$. Furthermore, the quantum-mechanical resonant momentum transfer from light to a two-level atom is also determined by the canonical momentum density [37,38]. Finally, a remarkable recent experiment [36], which realized quantum weak measurements of the local momentum of photons, also measured $\mathbf{p}^O$ [19]. Thus, the spin momentum $\mathbf{p}^S$ turns out to be indeed 'virtual', i.e., non-observable for weak-interaction measurements.

Note that we considered an 'ideal' particle with equal electric and magnetic polarizabilities. In reality, local light–matter interactions usually have electric character, and $|\alpha_e| \gg |\alpha_m|$. This is because of the fundamental electric–magnetic (dual) asymmetry of matter, which breaks the intrinsic dual symmetry of the free-space Maxwell equations [33,39,40]. In this case, the particle will 'measure' only the electric parts of the momentum and spin densities (5): $\mathbf{F} \simeq \gamma^{-1}\left[\frac{1}{2\omega}\text{Re}(\alpha_e)\nabla w_e + \text{Im}(\alpha_e)\mathbf{p}_e^O\right]$ and $\mathbf{T} \simeq \gamma^{-1}\text{Im}(\alpha_e)\mathbf{s}_e$ [25,33] (see Supplementary Note 3). The electric and magnetic contributions to the local dynamical characteristics of light are equivalent in paraxial propagating fields [22]: $\mathbf{p}_e^O \simeq \mathbf{p}_m^O$, $\mathbf{s}_e \simeq \mathbf{s}_m$, but they can differ significantly in other cases.

**Extraordinary momentum and spin in a single evanescent wave.** We are now in a position to consider the main subject of the present study: evanescent waves. A single evanescent wave propagating along the $z$-axis and decaying in the $x > 0$ half-space can formally be obtained via a rotation of the propagating plane wave (1) by an imaginary angle $i\vartheta$ about the $y$-axis [41]. In doing so, we obtain the electric evanescent-wave field:

$$\mathbf{E} = \frac{A}{\sqrt{1+|m|^2}} \left( \hat{\mathbf{x}} + m\frac{k}{k_z}\hat{\mathbf{y}} - i\frac{\kappa}{k_z}\hat{\mathbf{z}} \right) \exp(ik_z z - \kappa x). \qquad (7)$$

Here $k_z = k\cosh\vartheta > k$ is the longitudinal wave number, whereas $\kappa = k\sinh\vartheta$ is the exponential-decay rate, so that the complex wave vector is $\mathbf{k} = k_z\hat{\mathbf{z}} + i\kappa\hat{\mathbf{x}}$. Substituting field (7), with the corresponding magnetic wave field (see Supplementary Note 2 and Supplementary Figure 2), into equations (3) and (5), we calculate the canonical-momentum, spin-momentum, and spin-AM densities in the evanescent wave:

$$\mathbf{p}^O = \frac{w}{\omega} k_z \hat{\mathbf{z}}, \quad \mathbf{p}^S = \frac{w}{\omega}\left(-\frac{\kappa^2}{k_z}\hat{\mathbf{z}} + \sigma\frac{\kappa k}{k_z}\hat{\mathbf{y}}\right), \qquad (8)$$

$$\mathbf{s} = \frac{w}{\omega}\left(\sigma\frac{k}{k_z}\hat{\mathbf{z}} + \frac{\kappa}{k_z}\hat{\mathbf{y}}\right), \qquad (9)$$

where $w = \gamma\omega|A|^2\exp(-2\kappa x)$ is the spatially-inhomogeneous energy density of the wave.

Equations (8) and (9) reveal remarkable peculiarities of the momentum and spin in evanescent waves and represent the key analytical point of our study. First, note that the evanescent wave (7) possesses longitudinal canonical momentum $p_z^O \propto k_z > k$, which exceeds the momentum of a plane wave with the same local intensity. Divided by the energy density $w$, this momentum yields the superluminal local group velocity in evanescent waves: $v_{gz} = ck_z/k > c$ [19,42]. Although the Poynting vector corresponds to subluminal propagation, $p_z = p_z^O + p_z^S \propto (k^2/k_z) < k$, it is the canonical momentum that represents the observable momentum density. In particular, the momentum transfer via the radiation force (6) $F_z \propto p_z^O$ to a dipole particle in the evanescent wave (7) will be larger than $k$ per photon [43]. Such 'super-momentum' transfer was observed by Huard and Imbert [37] in the resonant Doppler coupling with a moving atom. In terms of the quantum weak-measurements paradigm, the 'super-momentum' $p_z^O$ represents a weak value of the photon momentum with the post-selection in a 'forbidden' zone unreachable for propagating waves (e.g., beyond a totally-reflecting interface) [19].

However, what is much more intriguing in equations (8) and (9) is the presence of the transverse $y$-components of the momentum and spin in the wave (7) propagating solely within the $(x,z)$ plane. Moreover, here the momentum $p_y^S \propto \sigma(\kappa k/k_z)w$ becomes proportional to the helicity $\sigma$, while the spin $s_y \propto (\kappa/k_z)w$ turns out to be helicity-independent! This is in sharp contrast to propagating waves and photons, Eqs. (2).

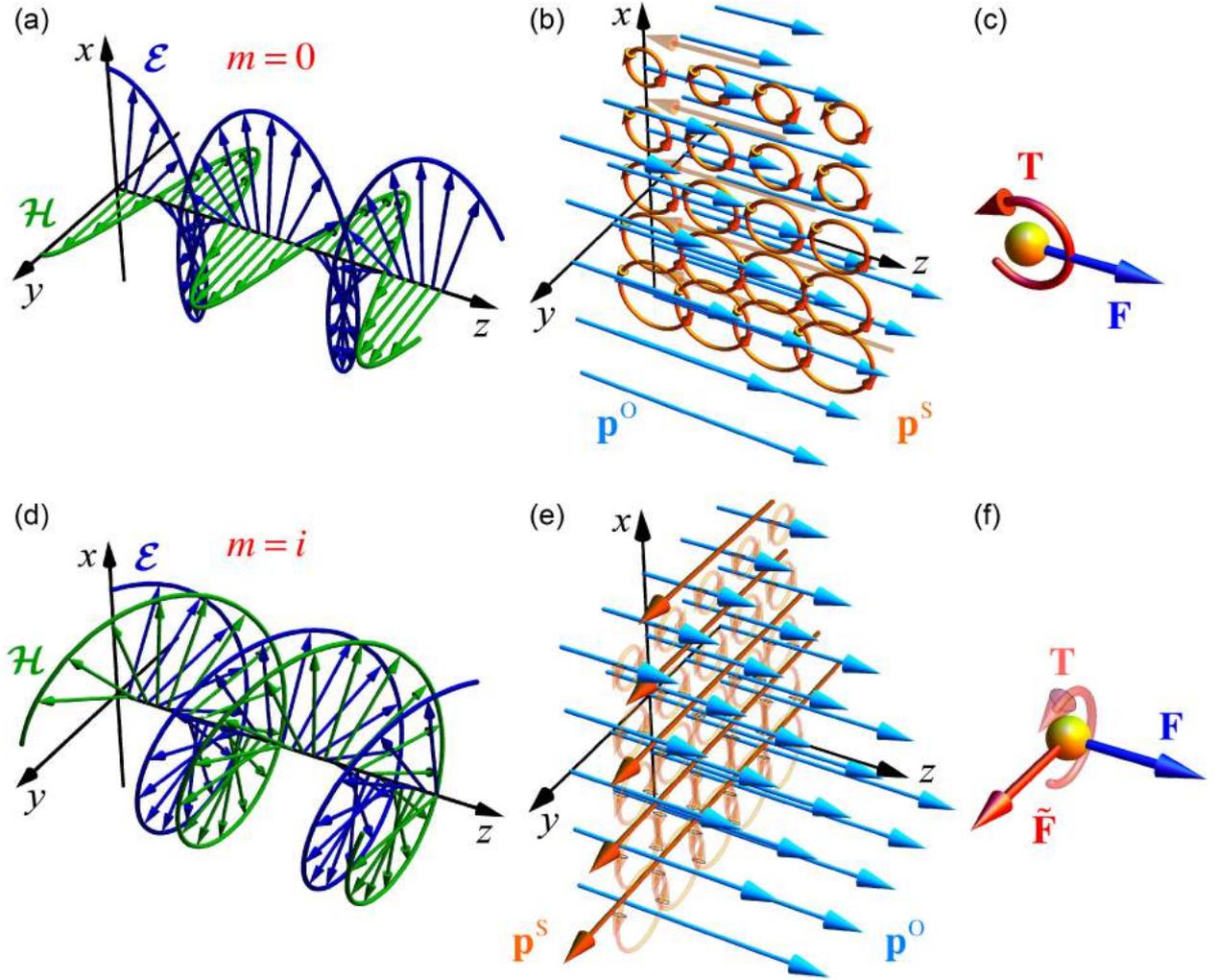

**Figure 2. Momentum and spin in linearly- and circularly-polarized evanescent plane waves.** The complex wave electric field is given by equation (7) with $m = \sigma = 0$ for the linear-polarization case **(a,b,c)** and $m = i$, $\sigma = 1$ for the circular-polarization case **(d,e,f)**. **(a,d)** The imaginary longitudinal components of the complex $\mathbf{E}(\mathbf{r})$ and $\mathbf{H}(\mathbf{r})$ fields (7) result in a cycloid-like projection of the instantaneous electric and magnetic field distributions, $\boldsymbol{\mathcal{E}}(\mathbf{r},t)$ and $\boldsymbol{\mathcal{H}}(\mathbf{r},t)$, onto the propagation $(x,z)$ plane (see Supplementary Note 2 and Supplementary Figure 2). As the wave propagates, the fields rotate in this plane even at linear polarizations. This rotation generates a transverse helicity-independent spin AM (8) $s_y \propto (\kappa/k_z)w$, represented in **(b)** by multiple loops of the spin momentum $\mathbf{p}^S$ in the $(x,z)$ plane. Due to the vertical inhomogeneity $w(x)$, these loops do not cancel each other, producing a backward spin momentum $p_z^S = \partial_x s_y / 2 \neq 0$ (semitransparent arrows). The circularly-polarized evanescent wave **(d)** also carries the usual longitudinal spin $s_z \propto \sigma w$ shown in **(e)** by multiple semitransparent loops of the spin momentum in the $(x,y)$ plane [cf. Fig. 1]. Due to the vertical inhomogeneity $w(x)$, these loops produce transverse helicity-dependent spin momentum (8) $p_y^S = -\partial_x s_z / 2 \propto \sigma(\kappa k/k_z)w$. The evanescent wave also possesses polarization-independent 'superluminal' orbital momentum (8) $\mathbf{p}^O \propto k_z \hat{\mathbf{z}} > k\hat{\mathbf{z}}$. **(c,f)** The orbital momentum and spin AM are locally transferred to the probe particle, thereby exerting: (i) an anomalously large radiation force $F_z \propto p_z^O \propto k_z > k$; (ii) the usual longitudinal helicity-dependent torque $T_z \propto s_z$; and (iii) the transverse helicity-independent torque $T_y \propto s_y$, Eqs. (6). The transverse spin momentum $p_y^S$ does not exert radiation

pressure in the dipole approximation (6), but does produce a helicity-dependent transverse force (10) $\tilde{F}_y \propto p_y^S$ (f) in higher-order interactions with larger Mie particles (see Fig. 3).

The transverse momentum and spin appear due to the two features of the evanescent field (7). The first one is the imaginary longitudinal component of the field polarization: $-i(\kappa/k_z)\hat{\mathbf{z}}$. This induces a rotation of the fields in the propagation $(x,z)$ plane [see Fig. 2a,d], and generates the spin $s_y \propto \text{Im}(E_z^* E_x + H_z^* H_x)$ independently of $\sigma$. Recently, we described such spin for surface plasmon-polaritons [42], and it was shown that the imaginary longitudinal field component plays an important role in optical coupling processes [44,45]. The second feature is the inhomogeneous intensity $w \propto \exp(-2\kappa x)$. This inhomogeneity destroys the cancellation of the spin-momentum loops in the $(x,y)$ plane, which results in the non-zero transverse Belinfante's spin momentum: $p_y^S = -\partial_x s_z/2 \neq 0$, see Fig. 2e. Note that although equations (8) and Figs. 2b,e show unidirectional spin momentum in the $x > 0$ half-space, the accurate consideration of the interface $x = 0$ and medium in the $x < 0$ half-space ensures the vanishing of the integral spin momentum, in agreement with $\int \mathbf{p}^S dV = 0$ [42].

**Relation to the Fedorov-Imbert controversy.** Here we should make a historical digression and note that an example of the transverse helicity-dependent momentum in an evanescent field was first found by F.I. Fedorov in 1955 [46]. Incidentally, this discovery caused a half-century-long controversy in the physics of light reflection and refraction. Analysing the total internal reflection of a polarized plane wave, Fedorov found a helicity-dependent transverse component of the Poynting vector in the transmitted evanescent field. Fedorov concluded that "the lateral energy flux should lead to a specific light pressure" and (by an analogy with the Goos–Hänchen effect) that "the reflected beam in the case of total reflection must be displaced in the lateral direction". Later, C. Imbert indeed observed such helicity-dependent beam shift experimentally [47], and the effect is now known as the Imbert–Fedorov transverse shift or spin Hall effect of light (for a review, see [48]).

The 50 years after Fedorov's finding brought about numerous controversies about this effect. Finally, only recently an accurate theoretical description was given [49,50], which was followed by a precise experiment [51] using quantum weak measurements (see also [52,53]). Remarkably, the current theory of the spin Hall effect of light is completely unrelated to evanescent waves and their transverse momentum. Indeed, the helicity-dependent beam shift arises from the interference of multiple plane waves in the beam, taking into account the geometric-phase effect, i.e., the spin-orbit interaction of light [48]. This shift occurs in partial reflection or refraction, focusing, scattering and other optical phenomena without any evanescent waves.

Thus, curiously, Fedorov predicted two fundamental helicity-dependent but unrelated phenomena (transverse momentum and beam shift) using a fictitious connection between them. Now, from equations (3) and (8), we can conclude that Fedorov's transverse momentum is an example of Belinfante's spin momentum, which does not transport energy and therefore cannot shift the field.

**Interaction of the evanescent wave with Mie particles.** The evanescent field (7) represents an exceptional configuration with a pure spin momentum without any orbital part in the transverse $y$-direction. This offers a unique opportunity to investigate this fundamental field-theory quantity *per se*. Equations (6) show that the spin momentum does not appear in the dipole interaction with small point-like particles. But does this result hold true for larger particles and higher-order interactions? To address this question, we examine the interaction of the evanescent optical field with finite-size Mie particles.

The Mie scattering theory provides an exact solution for the plane-wave interaction with an isotropic spherical particle. Using the Maxwell stress tensor, one can calculate the flux of the momentum and AM through a sphere enclosing the particle, and thereby determine the force and torque acting on the particle [54]. Recently, we developed and successfully tested an extension of the Mie theory (based on complex-angle rotation of the standard theory), which describes the scattering of the incident evanescent wave (6) [41] (see also [55,56]). Using this exact semi-analytical method, we calculate the radiation force and torque acting on the particle of radius $a$, complex permittivity $\varepsilon_p$, and permeability $\mu_p$, immersed in the evanescent field (7).

Figure 3a,c shows the schematic of the corresponding experiment using the total internal reflection at a glass prism. For this numerical experiment, we use parameters corresponding to real experiments manipulating particles with evanescent fields (e.g., [57–62]). Namely, we consider radiation with the wavelength $\lambda = 650$ nm, a gold particle ($\varepsilon_p = -12.2 + 3i$, $\mu_p = 1$) in water ($\varepsilon = 1.77$, $\mu = 1$), and near-critical total internal reflection (the angle of incidence is $\theta = 51° = \theta_c + 1.5°$) from the interface between heavy flint glass ($\varepsilon_1 = 3.06$, $\mu_1 = 1$) and water. Calculations of the corresponding wave fields and characteristics are given in the Supplementary Notes 2 and 3. The resulting force and torque components (normalized by $F_0 = a^2 |A_1|^2 / 4\pi$ and $T_0 = F_0 / k$, with $A_1$ being the amplitude of the incident wave in the glass), as functions of the dimensionless particle radius $ka$ are shown in Fig. 3b,d (see also Supplementary Figure 4 and Supplementary Table 1). These are the main numerical results of our work, which offer several new experiments for the detection of extraordinary spin and momentum properties of evanescent waves.

Figure 3b depicts the radiation torque components for right-hand and left-hand circularly-polarized waves ($m = \pm i$, $\sigma = \pm 1$). While the longitudinal torque $T_z$ from the usual spin $s_z$ flips with the sign of $\sigma$, the transverse torque $T_y$ is helicity-independent and present even in the linear-polarization $\text{Im}\, m = 0$ case. This confirms the presence of the transverse helicity-independent spin AM (8) $s_y$ in the evanescent field and its transfer to the particle. For small Rayleigh particles, $ka \ll 1$, the torque is described by the dipole approximation (6). Due to the strong dual (electric–magnetic) asymmetry of the gold, the torque appears mostly from the electric part of the spin (5): $\mathbf{T} \simeq \gamma^{-1} \text{Im}(\alpha_e) \mathbf{s}_e$, where the electric polarizability is proportional to the particle's volume: $\alpha_e \propto (ka)^3$, while the magnetic polarizability is small: $\alpha_m \propto (ka)^5 \simeq 0$ [63,64]. Therefore, the transverse torque is maximal for the TM-mode with $m = 0$ and $s_y = s_{ey}$, and minimal for the TE-mode with $m = \infty$ and $s_y = s_{my}$ (see Supplementary Notes 2 and 3).

The dual asymmetry results in another remarkable effect. Namely, for waves linearly diagonally-polarized at $\pm 45°$ ($m = \pm 1$), a vertical radiation torque $T_x$ appears, which is proportional to the degree of diagonal polarization $\chi = \dfrac{2\,\text{Re}\,m}{1 + |m|^2} \in [-1,1]$. This torque signals the presence of the vertical electric spin component:

$$s_{ex} = -s_{mx} = \chi \frac{\kappa k}{2 k_z^2} \frac{w}{\omega}, \tag{10}$$

which arises from the diagonal-electric-field rotation in the $(y,z)$-plane [see equation (7) and Supplementary Note 2]. Importantly, the total vertical spin vanishes in (9), because the electric and magnetic fields rotate in opposite directions: $s_x = s_{ex} + s_{mx} = 0$. Nonetheless, the dual-asymmetric gold particle unveils the electric vertical spin (10), as shown in Fig. 3b.

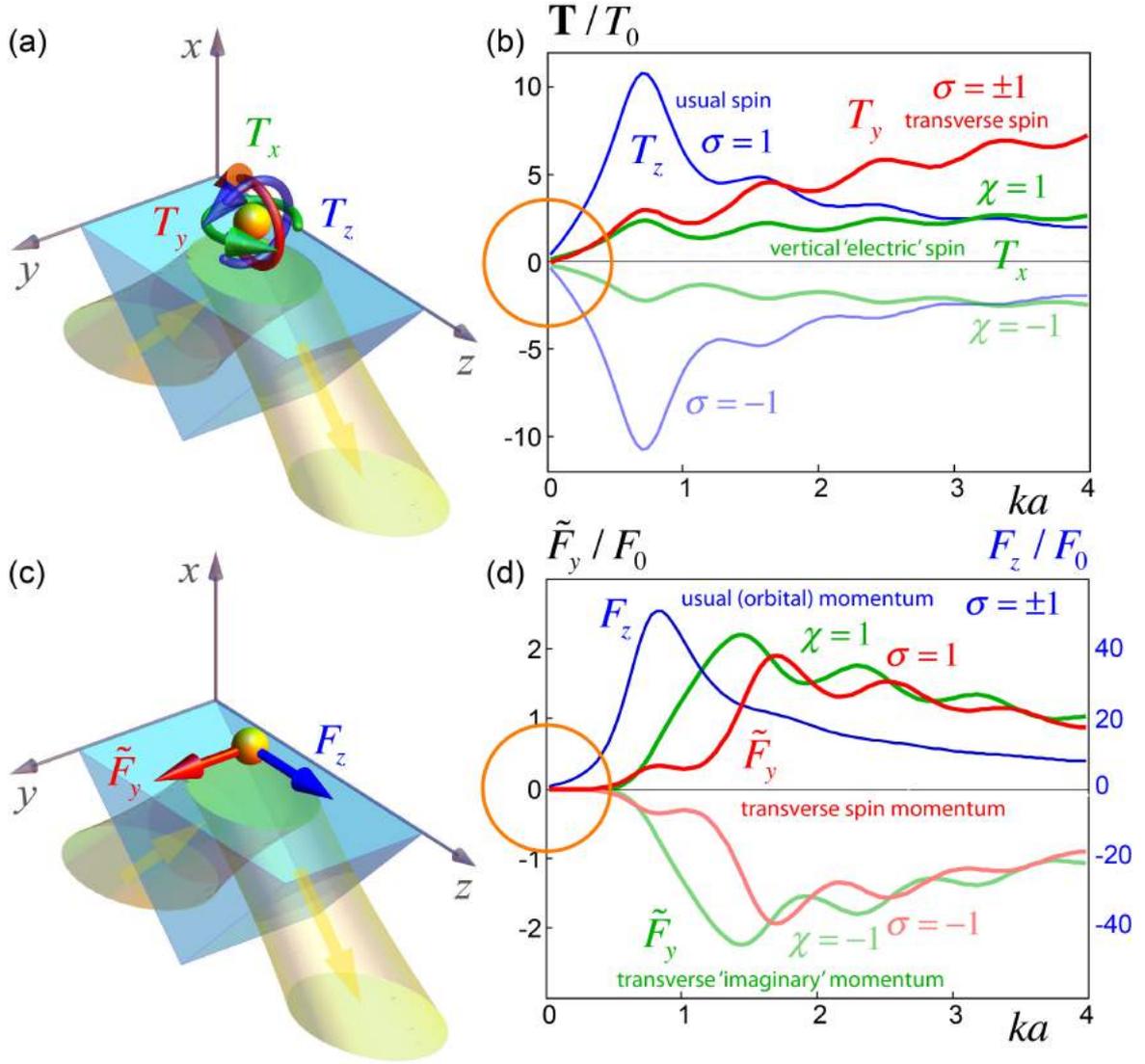

**Figure 3. Numerically-calculated forces and torques on a Mie particle in an evanescent field. (a,c)** Schematics of a proposed experiment (see also Supplementary Figure 3). A polarized propagating wave undergoes total internal reflection at the glass–water interface, thereby generating the evanescent wave (7) in water. A gold spherical particle of radius $a$ is placed in water on the glass surface, and its observable linear and spinning motion is proportional to the forces and torques exerted by the evanescent wave. **(b,d)** Normalized torque and force components for circular ($m = \pm i$, $\sigma = \pm 1$) and diagonal ($m = \chi = \pm 1$) polarizations versus the particle size $ka$ (the Rayleigh dipole region $ka \ll 1$ is indicated by the orange circles). **(b)** The longitudinal helicity-dependent torque $T_z \propto s_z \propto \sigma$ indicates the usual spin. The transverse torque is helicity-independent and signals the appearance of the transverse spin (9): $T_y \propto s_y \propto (\kappa/k_z)w$. The vertical torque for the diagonal polarizations $\chi = \pm 1$ appears because of the strong electric–magnetic (dual) asymmetry of the particle. It is caused by the non-zero electric part (10) of the zero-net vertical spin density: $T_x \propto s_{ex} \propto \chi(\kappa k/2k_z^2)w$, $s_{ex} = -s_{mx}$. **(d)** The orbital momentum density (8) produces mostly polarization-independent longitudinal radiation-pressure force: $F_z \propto p_z^O \propto k_z > k$. At the same time, the transverse force $\tilde{F}_y$ vanishes in the Rayleigh region, but becomes non-zero for larger Mie particles with $ka \sim 1$, see equation (11). It has the helicity-dependent part proportional to the transverse spin momentum (8): $\tilde{F}_y \propto p_y^S \propto \sigma(\kappa k/k_z)w$, and also the $\chi$-dependent part proportional to the 'imaginary' Poynting momentum (12) $\tilde{F}_y \propto \mathrm{Im}\,\tilde{p}_y \propto -\chi(\kappa k/k_z)w$.

Figure 3d shows the longitudinal and transverse components of the radiation force for circular ($m = \pm i$, $\sigma = \pm 1$) and diagonal ($m = \chi = \pm 1$) polarizations. The longitudinal force $F_z$ represents the radiation pressure (mostly polarization-independent) from the orbital momentum $p_z^O$. Akin to the torque, it exhibits an effective electric-dipole interaction $\mathbf{F} \simeq \gamma^{-1} \text{Im}(\alpha_e) \mathbf{p}_e^O$ in the Rayleigh regime $ka \ll 1$. The transverse force vanishes in this regime, $F_y = 0$, which confirms the 'virtual' character of the spin momentum. Nonetheless, a non-zero helicity-dependent transverse force arises for larger Mie particles with $ka \sim 1$. This force originates from the higher-order interaction between electric- and magnetic-induced dipoles, and in the quadratic dipole-dipole approximation it can be written as [63,64] (see Supplementary Note 3)

$$\tilde{\mathbf{F}} = \gamma^{-1} \frac{k^3}{3} \left[ -\text{Re}(\alpha_e \alpha_m^*) \text{Re}\,\tilde{\mathbf{p}} + \text{Im}(\alpha_e \alpha_m^*) \text{Im}\,\tilde{\mathbf{p}} \right]. \tag{11}$$

Here we introduced the complex Poynting momentum: $\tilde{\mathbf{p}} = \gamma k (\mathbf{E}^* \times \mathbf{H})$, with $\text{Re}\,\tilde{\mathbf{p}} = \mathbf{p} = \mathbf{p}^O + \mathbf{p}^S$ being the usual Poynting vector and $\text{Im}\,\tilde{\mathbf{p}} \propto \text{Im}(\mathbf{E}^* \times \mathbf{H})$ characterizing an alternating flow of the so-called 'stored energy' [2]. Alongside with the transverse $\sigma$-dependent spin momentum (8) $\text{Re}\,\tilde{p}_y = p_y^S \propto \sigma(\kappa k/k_z)w$, the evanescent wave (7) possesses a $\chi$-dependent transverse imaginary Poynting momentum (see Supplementary Note 3)

$$\text{Im}\,\tilde{p}_y = -\chi \frac{\kappa k}{k_z} \frac{w}{\omega}. \tag{12}$$

These two transverse momenta determine two contributions to the transverse dipole-dipole force (11) $\tilde{F}_y \propto -\text{Re}(\alpha_e \alpha_m^*) p_y^S + \text{Im}(\alpha_e \alpha_m^*) \text{Im}\,\tilde{p}_y$. In complete agreement with this, numerical calculations in Fig. 3d show $\sigma$-dependent and $\chi$-dependent transverse radiation forces on Mie particles with $ka \sim 1$, which vanish as $\tilde{F}_y \propto \alpha_e \alpha_m \propto (ka)^8$ at $ka \ll 1$. This proves the presence and observability of the transverse Belinfante's spin momentum in the evanescent optical field.

Detailed analysis and calculations of all torque and force components for all basic polarizations $m = 0, \infty, \pm i, \pm 1$ can be found in the Supplementary Note 3, Supplementary Figures 4 and 5, and Supplementary Table 1.

## Discussion

To summarize, we have found that a single evanescent electromagnetic wave offers a rich and highly non-trivial structure of the local momentum and spin distributions. In sharp contrast to standard photon properties, evanescent waves carry helicity-independent transverse spin and helicity-dependent transverse momentum. Moreover, the transverse momentum turns out to be a fundamental spin momentum introduced by Belinfante in field theory and first remarked in optics (as an unusual Poynting vector) by Fedorov. We have examined the measurements of the extraordinary spin and momentum in the evanescent field by analysing its interaction with a probe particle. Analytical evaluations and exact numerical simulations based on parameters of typical optical-manipulation experiments show that the transverse helicity-independent spin (and also the vertical electric spin for diagonal polarizations) can be detected straightforwardly via the radiation torque exerted on an absorbing small particle. At the same time, the Belinfante–Fedorov's spin momentum does not exert the standard optical pressure in the dipole approximation, which confirms its 'virtual' character (in contrast to Fedorov's interpretation). Nonetheless, it appears detectable (in contrast to the field-theory interpretation) via a helicity-dependent transverse optical force from the higher-order non-weak interaction with Mie particles. Thus, an exceptional evanescent-wave structure with pure spin transverse momentum offers a unique opportunity for the direct observation of this fundamental field-theory quantity, which was previously considered as 'virtual'.

In total, this work offers four novel experiments for the detection of extraordinary momentum and spin properties of a single evanescent wave (red and green curves in Fig. 3). These proposed experiments could detect the following optical torques and forces. First, the transverse helicity-independent torque $T_y$ indicating the transverse spin $s_y$ [Fig. 3b, equations (6) and (9)]. Second, the vertical diagonal-polarization-dependent torque $T_x$ exerted on a dual-asymmetric (e.g., electric-dipole) particle and caused by the vertical electric spin $s_{ex}$ [Fig. 3b and equation (10)]. Third, the transverse helicity-dependent force $\tilde{F}_y$ produced by the transverse spin momentum $p_y^S$ [Fig. 3d, equations (8) and (11)]. Fourth, the transverse diagonal-polarization-dependent force $\tilde{F}_y$, which is associated with the transverse imaginary Poynting vector $\text{Im}\,\tilde{p}_y$ [Fig. 3d, equations (11) and (12)].

These results add a distinct chapter in the physics of momentum and spin of classical and quantum fields, and offers a variety of non-trivial light-matter interaction effects involving evanescent fields.

## Acknowledgements

This work was partially supported by the RIKEN iTHES Project, MURI Center for Dynamic Magneto-Optics, JSPS-RFBR contract no. 12-02-92100, Grant-in-Aid for Scientific Research (S), MEXT Kakenhi on Quantum Cybernetics, and the JSPS via its FIRST program.

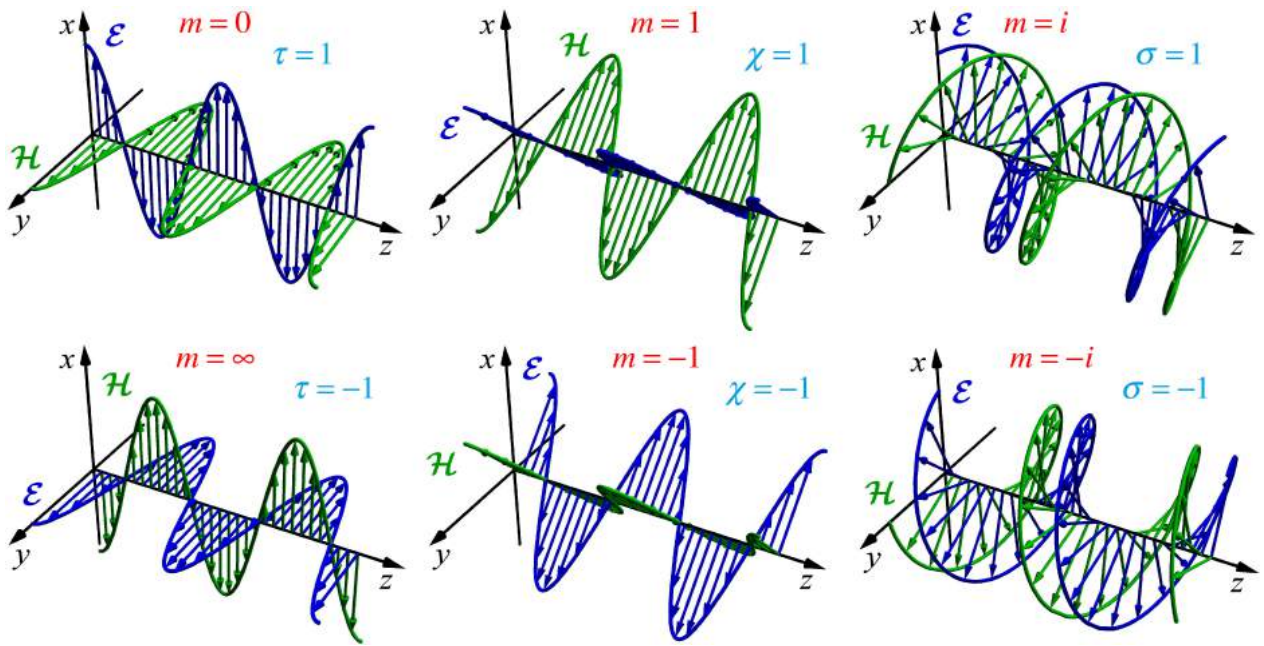

**Supplementary Figure 1.** Instantaneous distributions of the real electric and magnetic fields, $\mathcal{E}(\mathbf{r},t)$ and $\mathcal{H}(\mathbf{r},t)$, of the propagating plane wave (2.1). Six basic polarizations $\tau = \pm 1$, $\chi = \pm 1$, and $\sigma = \pm 1$ are shown and marked with the corresponding values of the complex polarization parameter $m$.

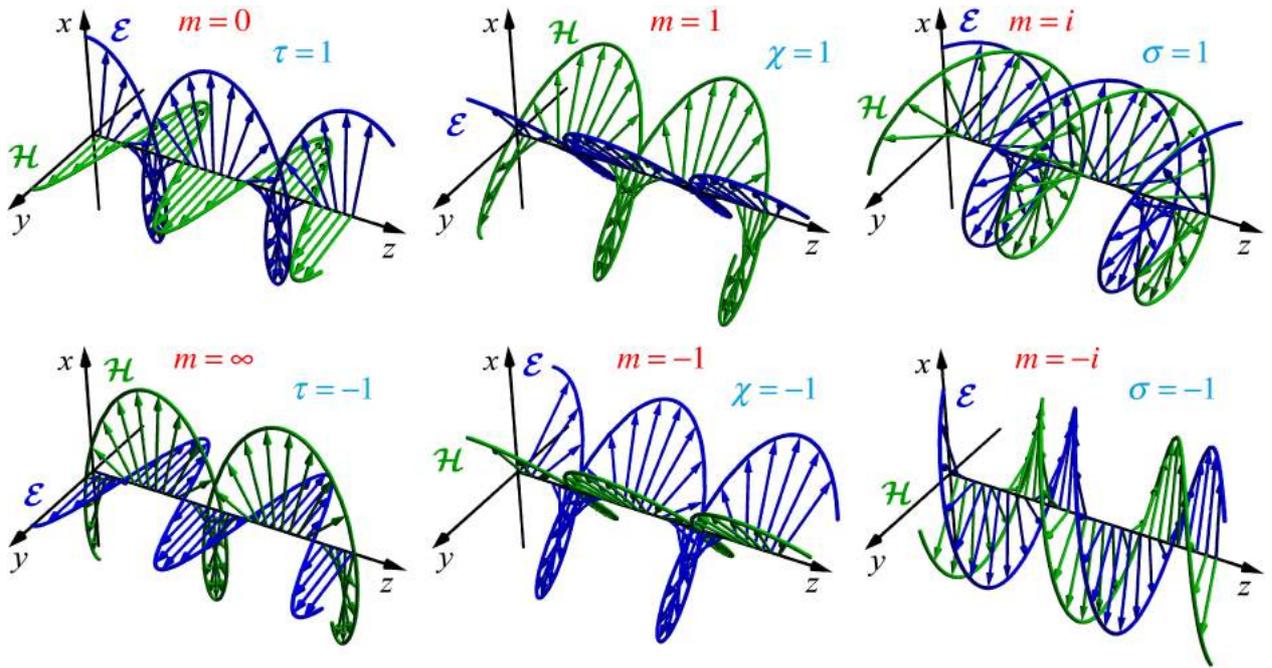

**Supplementary Figure 2.** Instantaneous distributions of the real electric and magnetic fields, $\mathcal{E}(\mathbf{r},t)$ and $\mathcal{H}(\mathbf{r},t)$, of the evanescent plane wave (2.4) and (2.5) [cf. the propagating plane wave in Supplementary Figure 1]. Six basic polarizations $\tau = \pm 1$, $\chi = \pm 1$, and $\sigma = \pm 1$ are shown and marked with the corresponding values of the complex polarization parameter $m$.

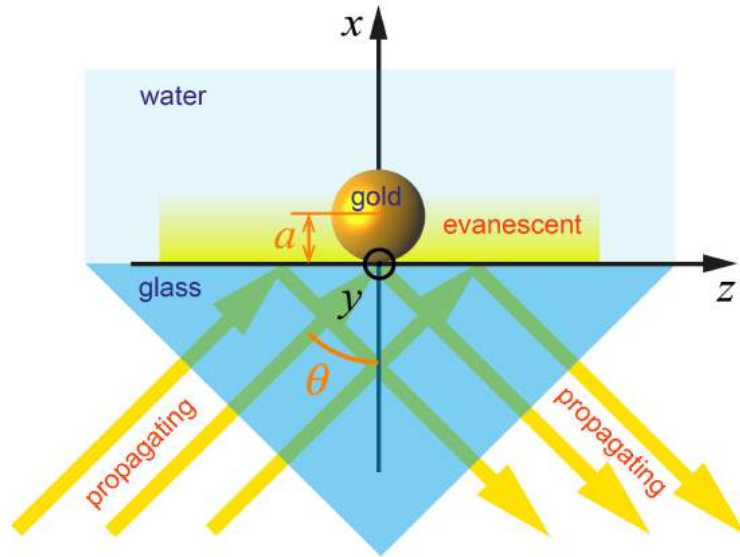

**Supplementary Figure 3.** Schematic of a proposed experiment observing the mechanical action of the evanescent field on a probe particle. The evanescent field is generated via the total internal reflection of the polarized light at the glass–water interface $x=0$. A spherical gold particle of radius $a$ is immersed in water on the surface of the glass. The radiation forces and torques cause linear and spinning motion of the particle, thereby, measuring the momentum and spin AM transfer from the evanescent field to the particle.

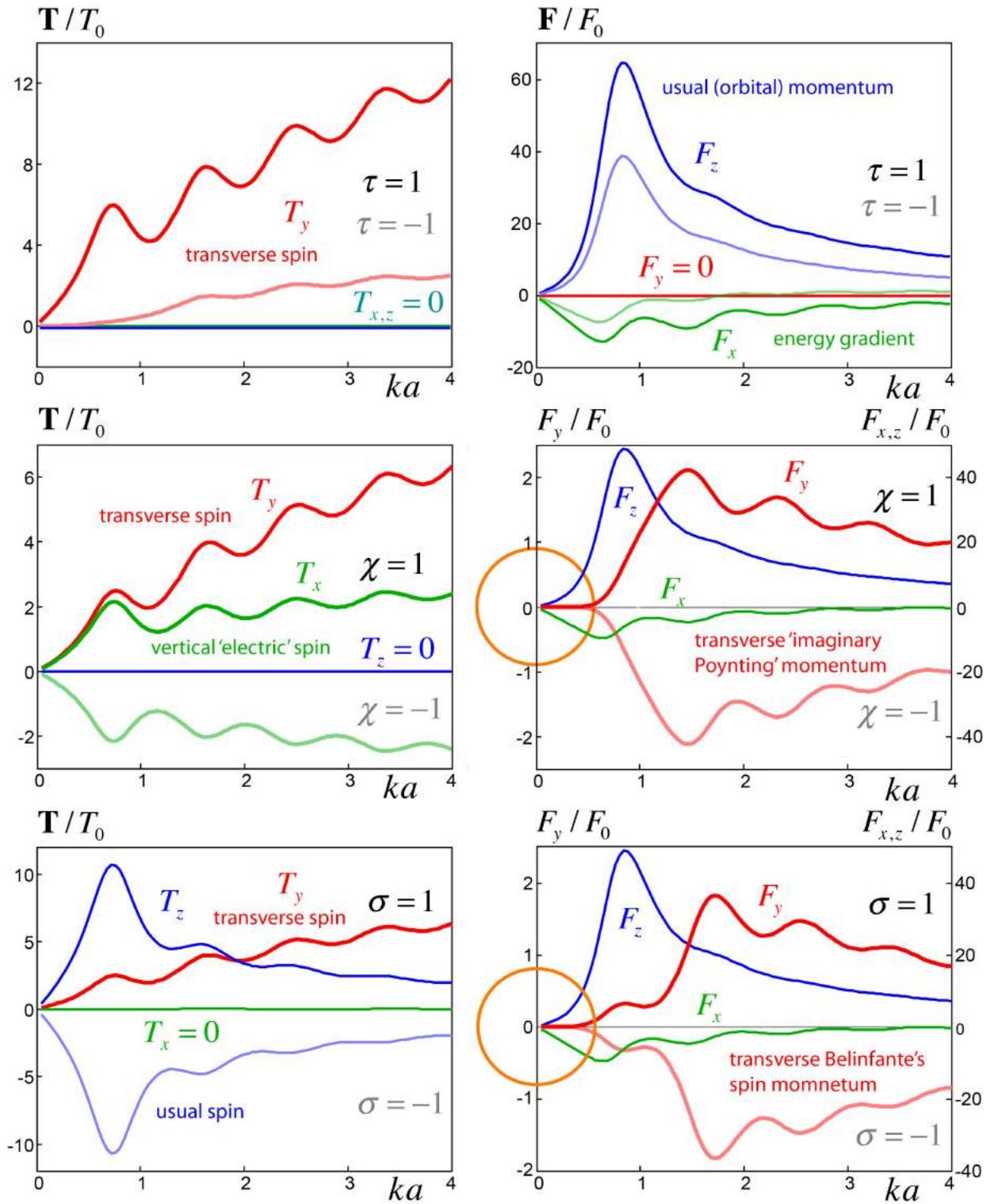

**Supplementary Figure 4.** Radiation forces and torques versus the particle size $ka$, numerically calculated for a gold Mie particle in the setup Supplementary Figure 3 with parameters (3.20). All components of the forces and torques are shown for six basic polarizations $\tau = \pm 1$, $\chi = \pm 1$, and $\sigma = \pm 1$ of the evanescent field. In addition to the known radiation-pressure longitudinal force, vertical gradient force, and longitudinal helicity-dependent torque, the following extraordinary forces and torques appear. The $\sigma$-independent torque $T_y$ indicates the transverse helicity-independent spin in the evanescent wave. The vertical $\chi$-dependent torque $T_x$ reveals the presence of the vertical electric spin in the diagonally-polarized

evanescent waves. Finally, the $\sigma$- and $\chi$-dependent transverse forces $F_y$ unveil the presence of the transverse Belinfante's spin momentum and 'imaginary' transverse Poynting vector (3.14). All the forces and torques and their correspondence to the field momenta and spins are summarized in Supplementary Table 1.

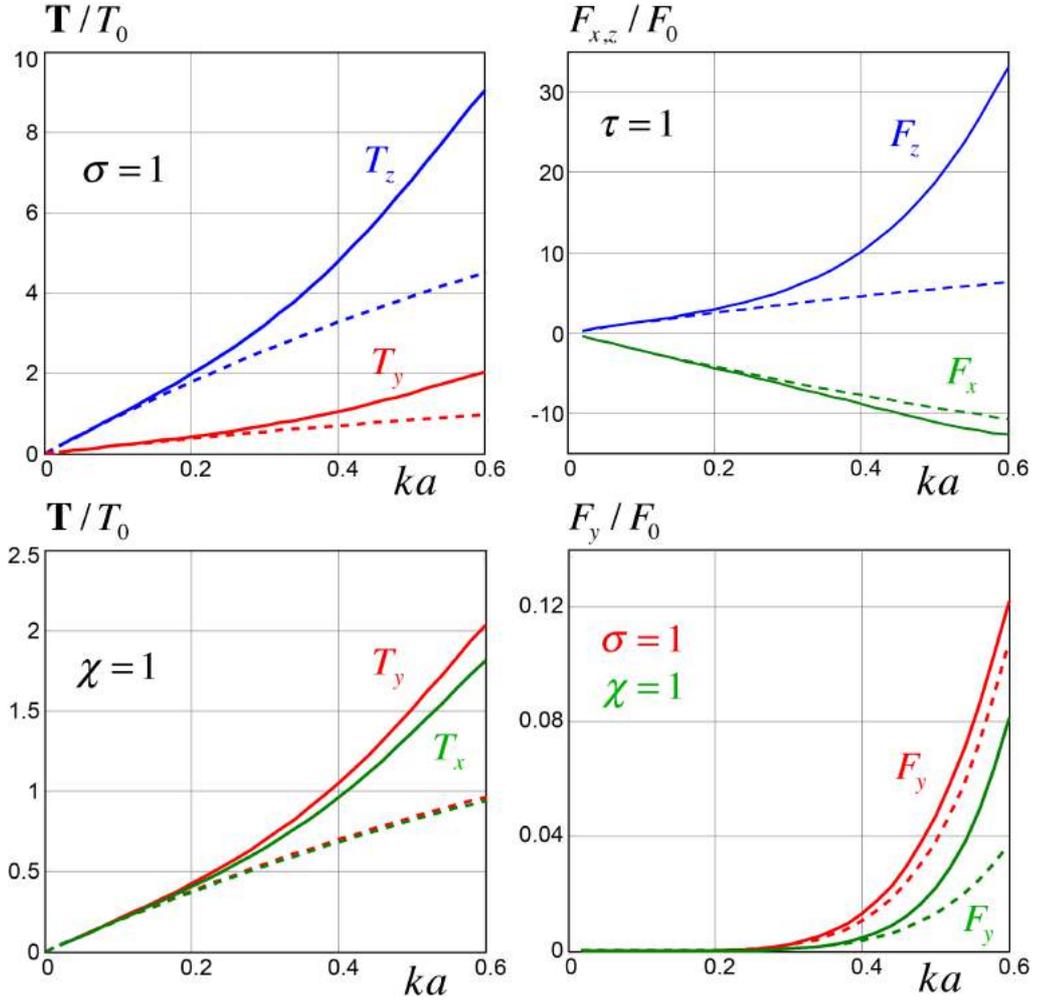

**Supplementary Figure 5.** Comparison between the numerical calculations (solid lines) and analytical dipole and dipole-dipole coupling approximations (dashed lines) for the radiation forces and torques on a small gold particle with $ka \ll 1$. Numerical calculations represent the exact Mie theory (the same as in Supplementary Figure 4), whereas the dipole and dipole-dipole approximations are described by Eqs. (3.6)–(3.15). One can see that the analytical equations describe the leading orders of the forces and torques in the Rayleigh-particle region $ka \ll 1$, but the exact forces and torques become significantly larger in the strong-coupling Mie region with $ka \sim 1$.

| Field characteristics | Action on a small particle |
|---|---|
| Longitudinal canonical momentum $$p_z^{\mathrm{O}} \propto k_z w$$ | Longitudinal $\tau$-dependent electric-dipole radiation-pressure force $$F_z \propto p_{\mathrm{e}z}^{\mathrm{O}} \propto \left(1 + \tau \frac{\kappa^2}{k_z^2}\right) p_z^{\mathrm{O}}$$ |
| Vertical imaginary canonical momentum $$\mathrm{Im}\,\tilde{p}_x^{\mathrm{O}} \propto -\nabla_x w / 2$$ | Vertical $\tau$-dependent electric-dipole gradient force $$F_x \propto -\mathrm{Im}\,\tilde{p}_{\mathrm{e}x}^{\mathrm{O}} \propto -\left(1 + \tau \frac{\kappa^2}{k_z^2}\right)\mathrm{Im}\,\tilde{p}_x^{\mathrm{O}}$$ |
| Transverse helicity-dependent spin (Poynting) momentum $$p_y^{\mathrm{S}} = p_y \propto \sigma(\kappa k / k_z) w$$ | Transverse $\sigma$-dependent part of the dipole-dipole force (3.12) $$\tilde{F}_y \propto -p_y^{\mathrm{S}}$$ |
| Transverse imaginary Poynting momentum at diagonal polarizations $$\mathrm{Im}\,\tilde{p}_y \propto -\chi(\kappa k / k_z) w$$ | Transverse $\chi$-dependent part of the dipole-dipole force (3.12) $$\tilde{F}_y \propto \mathrm{Im}\,\tilde{p}_y$$ |
| Longitudinal helicity-dependent spin $$s_z \propto \sigma(k / k_z) w$$ | Longitudinal $\sigma$-dependent electric-dipole torque $$T_z \propto s_{\mathrm{e}z} \propto s_z$$ |
| Transverse polarization-independent spin $$s_y \propto (\kappa / k_z) w$$ | Transverse $\tau$-dependent electric-dipole torque $$T_y \propto s_{\mathrm{e}y} \propto (1 + \tau) s_y$$ |
| Vertical electric and magnetic spins at diagonal polarizations $$s_{\mathrm{e}x} = -s_{\mathrm{m}x} \propto \chi(\kappa k / k_z^2) w$$ | Vertical $\chi$-dependent electric-dipole torque $$T_x \propto s_{\mathrm{e}x}$$ |

**Supplementary Table 1.** Four distinct momenta and three spins in a polarized evanescent wave versus their observable manifestations in the forces and torques on a small particle with $ka \ll 1$. One can trace the exact correspondence of these polarization-dependent forces and torques to the exact results (for $ka \sim 1$) shown in Supplementary Figure 4.

## Supplementary Note 1. Harmonic Maxwell fields and their characteristics

We consider Maxwell equations for monochromatic electromagnetic fields in a uniform non-dispersive medium with permittivity $\varepsilon$ and permeability $\mu$:

$$\nabla \cdot \mathbf{H} = \nabla \cdot \mathbf{E} = 0, \quad -i\frac{\omega}{c}\varepsilon \mathbf{E} = \nabla \times \mathbf{H}, \quad -i\frac{\omega}{c}\mu \mathbf{H} = -\nabla \times \mathbf{E}. \tag{1.1}$$

Here $\omega$ is the frequency, $c$ is the speed of light in vacuum, $\mathbf{E}(\mathbf{r})$ and $\mathbf{H}(\mathbf{r})$ are the complex electric and magnetic field amplitudes, whereas the real fields are given by $\mathcal{E}(\mathbf{r},t) = \text{Re}\left[\mathbf{E}(\mathbf{r})e^{-i\omega t}\right]$ and $\mathcal{H}(\mathbf{r},t) = \text{Re}\left[\mathbf{H}(\mathbf{r})e^{-i\omega t}\right]$, and throughout the paper we use Gaussian units.

The time-averaged energy density and Poynting momentum density of the monochromatic field are known to be [1]

$$w = \frac{g}{4}\left(\varepsilon|\mathbf{E}|^2 + \mu|\mathbf{H}|^2\right), \quad \mathbf{p} = \frac{g}{2c}\text{Re}\left(\mathbf{E}^* \times \mathbf{H}\right), \tag{1.2}$$

where $g = 1/4\pi$ is the Gaussian-units coefficient. Although the Poynting vector $\mathbf{p}$ is usually considered as a meaningful momentum density of the field, in fact it represents a sum of two terms with quite different physical meanings. Using Maxwell equations (1.1), one can write it as $\mathbf{p} = \mathbf{p}^O + \mathbf{p}^S$ [2–5]:

$$\mathbf{p}^O = \frac{g}{4\omega}\text{Im}\left[\mu^{-1}\mathbf{E}^*\cdot(\nabla)\mathbf{E} + \varepsilon^{-1}\mathbf{H}^*\cdot(\nabla)\mathbf{H}\right], \tag{1.3}$$

$$\mathbf{p}^S = \frac{g}{8\omega}\nabla \times \text{Im}\left[\mu^{-1}\left(\mathbf{E}^* \times \mathbf{E}\right) + \varepsilon^{-1}\left(\mathbf{H}^* \times \mathbf{H}\right)\right]. \tag{1.4}$$

Here $\mathbf{p}^O$ is the *canonical* or *orbital momentum density*, which is responsible for the energy transport and radiation pressure, whereas $\mathbf{p}^S$ is the *spin momentum density*, which does not transport energy but generates the spin angular momentum (AM) of light [3,6–10]. (The spin momentum was introduced by Belinfante [7] in field theory for the explanation of spin of quantum particles and symmetrisation of the energy–momentum tensor.) It is the canonical momentum density $\mathbf{p}^O$ that represents the observable momentum of light [2,3,11–15]; it characterizes the local wave vector of the field (multiplied by the intensity), which is mostly independent of the polarization. At the same time, the spin momentum density $\mathbf{p}^S$ is a virtual solenoidal current, given by the curl of the *spin AM density*, $\mathbf{p}^S = \nabla \times \mathbf{s}/2$:

$$\mathbf{s} = \frac{g}{4\omega}\text{Im}\left[\mu^{-1}\left(\mathbf{E}^* \times \mathbf{E}\right) + \varepsilon^{-1}\left(\mathbf{H}^* \times \mathbf{H}\right)\right]. \tag{1.5}$$

As it can be seen from their names, the orbital and spin momentum densities are responsible for the generation of the orbital and spin AM of light. Namely, the *orbital AM density* is defined in a straightforward way as $\mathbf{l} = \mathbf{r} \times \mathbf{p}^O$, and this is an extrinsic origin-dependent quantity. At the same time, the spin AM density $\mathbf{s}$, Eq. (1.5), is intrinsic (origin-independent). Nonetheless, its integral value is determined by the circulation of the spin momentum (1.4): $\mathbf{S} = \int \mathbf{s}\,dV = \int \mathbf{r} \times \mathbf{p}^S\,dV$, where integration by parts should be performed [3–10]. Thus, the spin momentum is similar to the boundary magnetization current or topological quantum-Hall-state current in solid-state systems, whereas the spin AM is analogous to the bulk magnetization in such systems.

In addition to the above dynamical characteristics of the field, there is one more fundamental quantity, namely, the *helicity density*. Recently, it caused considerable attention [5,16,17] in connection with the fundamental dual 'electric–magnetic' symmetry of Maxwell equations [18–20] and optical interaction with chiral particles [21–24]. The time-averaged helicity density of the monochromatic Maxwell field can be written as [5]

$$h = -\frac{g}{2\omega} \text{Im}\left(\mathbf{E}^* \cdot \mathbf{H}\right). \tag{1.6}$$

The helicity characterizes the difference between the number of right-hand and left-hand circularly-polarized photons.

In free space, $\varepsilon = \mu = 1$, bilinear quantities (1.2)–(1.6) allow a convenient quantum-like representation in terms of the energy, momentum, and spin operators [2,5]. To show this, we introduce the local state vector of the field:

$$\vec{\psi}(\mathbf{r}) = \sqrt{\frac{g}{4\omega}} \begin{pmatrix} \mathbf{E}(\mathbf{r}) \\ \mathbf{H}(\mathbf{r}) \end{pmatrix}. \tag{1.7}$$

This is formally a vector in $\mathbb{C}^3 \otimes \mathbb{C}^2 \otimes L^2$ space, where the 'dual' $\mathbb{C}^2$ space is associated with the electric and magnetic degrees of freedom. (Rigorously speaking, monochromatic fields are not square-integrable functions, but this does not affect our local analysis.) Using the state vector (1.7), the energy, canonical momentum, spin AM, and helicity densities can be written as 'local expectation values' of the corresponding operators:

$$w = \vec{\psi}^\dagger \cdot (\omega) \vec{\psi}, \tag{1.8}$$

$$\mathbf{p}^O = \text{Re}\left[\vec{\psi}^\dagger \cdot (\hat{\mathbf{p}}) \vec{\psi}\right], \tag{1.9}$$

$$\mathbf{s} = \vec{\psi}^\dagger \cdot (\hat{\mathbf{S}}) \vec{\psi}, \tag{1.10}$$

$$h = \vec{\psi}^\dagger \cdot (-\hat{\sigma}_2) \vec{\psi} = \vec{\psi}^\dagger \cdot \left(\frac{\hat{\mathbf{p}} \cdot \hat{\mathbf{S}}}{p}\right) \vec{\psi}. \tag{1.11}$$

Here $\hat{\mathbf{p}} = -i\nabla$ is the canonical momentum operator in $L^2$ (we use units $\hbar = 1$). The spin operator $\hat{\mathbf{S}}$ in $\mathbb{C}^3$ is given by spin-1 matrices:

$$\hat{S}_x = -i\begin{pmatrix} 0 & 0 & 0 \\ 0 & 0 & 1 \\ 0 & -1 & 0 \end{pmatrix}, \quad \hat{S}_y = -i\begin{pmatrix} 0 & 0 & -1 \\ 0 & 0 & 0 \\ 1 & 0 & 0 \end{pmatrix}, \quad \hat{S}_z = -i\begin{pmatrix} 0 & 1 & 0 \\ -1 & 0 & 0 \\ 0 & 0 & 0 \end{pmatrix}, \tag{1.12}$$

(which act on the electric- and magnetic-field components as $\mathbf{E}^* \cdot (\hat{\mathbf{S}}) \mathbf{E} = \text{Im}(\mathbf{E}^* \times \mathbf{E})$ and $\mathbf{H}^* \cdot (\hat{\mathbf{S}}) \mathbf{H} = \text{Im}(\mathbf{H}^* \times \mathbf{H})$). Finally, the 'dual' operator in $\mathbb{C}^2$

$$\hat{\sigma}_2 = \begin{pmatrix} 0 & -i \\ i & 0 \end{pmatrix} \tag{1.13}$$

mixes the electric and magnetic subspaces. The eigenmodes of this operator, $\hat{\sigma}_2 \vec{\psi} = \sigma \vec{\psi}$, with $\sigma = \pm 1$, are the fields with well-defined helicity: $\mathbf{H} = -i\sigma \mathbf{E}$. The last equality in Eq. (1.11) with $p = \omega/c$ represents a form of the last two Maxwell equations (1.1) in vacuum. This reveals the connection to the quantum-mechanical helicity as the spin projection onto the momentum direction. The Poynting momentum (1.2) can also be written in a quantum-like form, and it is characterized by a mixed energy–helicity–spin operator:

$$\mathbf{p} = \vec{\psi}^\dagger \cdot \left( -\omega \hat{\sigma}_2 \otimes \hat{\mathbf{S}} \right) \vec{\psi} \,. \tag{1.14}$$

Note that the 'local expectation values' of the quantum operators, Eqs. (1.8)–(1.11) and (1.14), can be interpreted in terms of *quantum weak measurements* [25–27]. For any operator $\hat{\mathbf{O}}$, the corresponding local density $\mathbf{O}(\mathbf{r})$ is proportional to the real part of the complex *non-normalized weak value* $\tilde{\mathbf{O}}(\mathbf{r})$ with the post-selection in the coordinate eigenstate [2,5,11,12]:

$$\tilde{\mathbf{O}}(\mathbf{r}) = \vec{\psi}^\dagger(\mathbf{r}) \cdot (\hat{\mathbf{O}}) \vec{\psi}(\mathbf{r}) = \langle \vec{\psi} | \mathbf{r} \rangle \langle \mathbf{r} | \hat{\mathbf{O}} | \vec{\psi} \rangle. \tag{1.15}$$

(As usual, 'bra' and 'ket' notations are used for the inner product in the $L^2$ Hilbert space.) In particular, the canonical momentum density (1.9) is the real part of the *complex* canonical momentum, or 'weak momentum':

$$\tilde{\mathbf{p}}^O(\mathbf{r}) = \vec{\psi}^\dagger(\mathbf{r}) \cdot (\hat{\mathbf{p}}) \vec{\psi}(\mathbf{r}) = \langle \vec{\psi} | \mathbf{r} \rangle \langle \mathbf{r} | \hat{\mathbf{p}} | \vec{\psi} \rangle = \mathbf{p}^O(\mathbf{r}) - i \frac{1}{2\omega} \nabla w(\mathbf{r}). \tag{1.16}$$

Indeed, a recent remarkable experiment [12], which realized quantum weak measurements of the local momentum of photons, in fact measured $\mathbf{p}^O(\mathbf{r})$ (with the imaginary part of the weak value (1.16) being also observable) [11].

Importantly, almost all meaningful dynamical characteristics (1.2)–(1.5), except for the helicity (1.6), naturally represent a sum of the electric- and magnetic-field contributions:

$$w = w_e + w_m, \quad \mathbf{p}^O = \mathbf{p}_e^O + \mathbf{p}_m^O, \quad \mathbf{p}^S = \mathbf{p}_e^S + \mathbf{p}_m^S, \quad \mathbf{s} = \mathbf{s}_e + \mathbf{s}_m. \tag{1.17}$$

The symmetry between the electric and magnetic contributions reflects the dual symmetry of the free-space Maxwell equations and fields. At the same time, matter is typically strongly dual-asymmetric (since it is built using electric but not magnetic charges), so that the electric and magnetic parts of the field characteristics can play very different roles in light–matter interactions (including measurement processes). The helicity (1.6) mixes electric and magnetic fields because it represents the generator of the dual transformations of Maxwell equations [5,16–20].

## Supplementary Note 2. Application to evanescent wave fields

**Evanescent wave fields.** Consider first a polarized electromagnetic plane wave propagating along the $z$-axis in a medium with permittivity $\varepsilon$ and permeability $\mu$. The complex electric and magnetic fields can be written in Cartesian coordinates as

$$\mathbf{E}(\mathbf{r}) = \frac{A\sqrt{\mu}}{\sqrt{1+|m|^2}} \begin{pmatrix} 1 \\ m \\ 0 \end{pmatrix} \exp(ikz), \quad \mathbf{H}(\mathbf{r}) = \frac{A\sqrt{\varepsilon}}{\sqrt{1+|m|^2}} \begin{pmatrix} -m \\ 1 \\ 0 \end{pmatrix} \exp(ikz), \qquad (2.1)$$

where $A$ is the wave amplitude, $k = n\omega/c$ is the wavenumber, $n = \sqrt{\varepsilon\mu}$ is the refraction index of the medium, and the complex number $m$ characterizes the polarization of the wave. Namely, $m=0$ and $m=\infty$ correspond to the $x$ (TM) and $y$ (TE) linear polarizations; $m = \pm 1$ correspond to the diagonal and anti-diagonal linear polarizations at $\pm 45°$, and $m = \pm i$ describe the right- and left-hand circular polarizations. The degrees of the TE–TM, diagonal, and circular polarizations are described by the corresponding normalized Stokes parameters:

$$\tau = \frac{1-|m|^2}{1+|m|^2}, \quad \chi = \frac{2\operatorname{Re} m}{1+|m|^2}, \quad \sigma = \frac{2\operatorname{Im} m}{1+|m|^2}. \qquad (2.2)$$

Here $\tau^2 + \chi^2 + \sigma^2 = 1$, and the third Stokes parameter $\sigma$ determines the helicity (1.6) of the wave. Supplementary Figure 1 shows instantaneous distributions of the real electric and magnetic fields, $\boldsymbol{\mathcal{E}}(\mathbf{r},t)$ and $\boldsymbol{\mathcal{H}}(\mathbf{r},t)$, in the propagating wave (2.1) for six basic polarizations: $\tau = \pm 1$, $\chi = \pm 1$, and $\sigma = \pm 1$.

An evanescent plane wave propagating along the $z$-axis and decaying along the $x$-axis can be obtained from the plane wave (2.1) via an imaginary-angle rotation about the transverse $y$-axis [28]. Such rotation is described by the transformation matrix

$$\hat{R}(i\vartheta) = \begin{pmatrix} \cos(i\vartheta) & 0 & \sin(i\vartheta) \\ 0 & 1 & 0 \\ -\sin(i\vartheta) & 0 & \cos(i\vartheta) \end{pmatrix} = \begin{pmatrix} \cosh\vartheta & 0 & i\sinh\vartheta \\ 0 & 1 & 0 \\ -i\sinh\vartheta & 0 & \cosh\vartheta \end{pmatrix}. \qquad (2.3)$$

Applying it to both vector components and spatial distribution of the fields (2.1), $\mathbf{E}(\mathbf{r}) \to \hat{R}(i\vartheta)\mathbf{E}[\hat{R}(-i\vartheta)\mathbf{r}]$, $\mathbf{H}(\mathbf{r}) \to \hat{R}(i\vartheta)\mathbf{H}[\hat{R}(-i\vartheta)\mathbf{r}]$, we derive the complex evanescent-wave fields [28]:

$$\mathbf{E}(\mathbf{r}) = \frac{A\sqrt{\mu}}{\sqrt{1+|m|^2}} \begin{pmatrix} 1 \\ mk/k_z \\ -i\kappa/k_z \end{pmatrix} \exp(ik_z z - \kappa x), \qquad (2.4)$$

$$\mathbf{H}(\mathbf{r}) = \frac{A\sqrt{\varepsilon}}{\sqrt{1+|m|^2}} \begin{pmatrix} -m \\ k/k_z \\ im\kappa/k_z \end{pmatrix} \exp(ik_z z - \kappa x). \qquad (2.5)$$

Here we introduced the propagation constant $k_z = k\cosh\vartheta > k$, the decay constant $\kappa = k\sinh\vartheta$, $k_z^2 - \kappa^2 = k^2$, and also renormalized the amplitude $A \to A/\cosh\vartheta$. One can readily verify that fields (2.4) and (2.5) satisfy Maxwell equations (1.1).

Supplementary Figure 2 shows the instantaneous distributions of the real electric and magnetic fields, $\mathcal{E}(\mathbf{r},t)$ and $\mathcal{H}(\mathbf{r},t)$, in the evanescent wave (2.4) and (2.5) for six basic polarizations: $\tau = \pm 1$, $\chi = \pm 1$, and $\sigma = \pm 1$. The main difference, compared to the propagating wave (Supplementary Figure 1), is the presence of the *imaginary longitudinal $z$-components* in the complex fields (2.4) and (2.5). These components result in rotations of the electric and magnetic fields in the propagation $(x,z)$ plane, which generate the transverse helicity-independent spin [see Eq. (2.14) below]. Furthermore, for diagonal polarizations $\chi = \pm 1$, the electric and magnetic fields also rotate in the $(y,z)$ plane, thereby generating the vertical electric and magnetic spins of the opposite signs [see Eq. (2.18) below].

The simplest way to generate the evanescent wave (2.4) and (2.5) in the $x > 0$ half-space is to use the total internal reflection at the $x = 0$ interface between a medium with parameters $\varepsilon_1$ and $\mu_1$ (e.g., glass), and the medium with parameters $\varepsilon$ and $\mu$ (e.g., water), such that $n_1 = \sqrt{\varepsilon_1 \mu_1} > n$. Let the incident plane wave of the type (2.1) with amplitude $A_1$, polarization $m_1$, and wave number $k_1 = n_1 \omega / c$ propagate in the glass at the angle $\theta > \theta_c \equiv \sin^{-1}(n/n_1)$ with respect to the $x$-axis, Supplementary Figure 3. Then, the transmitted wave in water ($x > 0$) will be the evanescent wave (2.4) and (2.5) with the parameters given by

$$A = \frac{k_z}{k}\sqrt{\frac{\mu_1}{\mu}} T A_1, \quad k_z = k\frac{n_1}{n}\sin\theta, \quad \kappa = k\sqrt{\left(\frac{n_1}{n}\right)^2 \sin^2\theta - 1}, \quad m = \frac{T_\perp}{T_\parallel} m_1. \tag{2.6}$$

Here

$$T_\parallel = -\frac{2k\sqrt{\varepsilon_1/\mu_1}\cos\theta_1}{k\sqrt{\varepsilon/\mu}\cos\theta_1 + i\kappa\sqrt{\varepsilon_1/\mu_1}}, \quad T_\perp = -\frac{2k\sqrt{\varepsilon_1/\mu_1}\cos\theta_1}{k\sqrt{\varepsilon_1/\mu_1}\cos\theta_1 + i\kappa\sqrt{\varepsilon/\mu}} \tag{2.7}$$

are the Fresnel transmission coefficients for the TM and TE polarization components [1], whereas $T = \dfrac{\sqrt{|T_\parallel|^2 + |m_1|^2 |T_\perp|^2}}{\sqrt{1 + |m_1|^2}} \exp\left[i \arg T_\parallel\right]$. In the case of near-critical incidence, $0 < \theta - \theta_c \ll 1$, the polarization of the transmitted evanescent wave can be approximated as $m \simeq \sqrt{\varepsilon \mu_1 / \varepsilon_1 \mu}\, m_1$, which shows a way for generating the evanescent wave of any desired polarization $m$ by preparing the corresponding polarization $m_1$ of the incident wave.

**Local characteristics of the evanescent field.** Now we calculate the local dynamical characteristics of the evanescent wave (2.4) and (2.5). As a reference point, we first find the energy, momentum, spin, and helicity densities in the propagating plane wave (2.1). In this case, equations (1.2)–(1.6) result in

$$w = \gamma I n^2 \omega, \tag{2.8}$$

$$\mathbf{p}^O = \mathbf{p} = \gamma I k \hat{\mathbf{z}}, \quad \mathbf{p}^S = \mathbf{0}, \tag{2.9}$$

$$\mathbf{s} = \gamma I \sigma \hat{\mathbf{z}}, \quad h = \gamma I n \sigma, \tag{2.10}$$

where $\gamma = g/2\omega$, $I = |A|^2$ is the wave intensity, and $\hat{\mathbf{x}}, \hat{\mathbf{y}}, \hat{\mathbf{z}}$ denote the unit vectors of the corresponding axes. In vacuum ($n = 1$), equations (2.8)–(2.10) yield

$$\mathbf{p} = \frac{w}{\omega} k \hat{\mathbf{z}}, \quad \mathbf{s} = \frac{w}{\omega} \sigma \hat{\mathbf{z}}, \quad h = \frac{w}{\omega} \sigma, \tag{2.11}$$

in agreement with the quantum-mechanical picture of photons.

Now, substituting fields (2.4) and (2.5) into equations (1.2)–(1.6), we derive the energy, momentum, spin, and helicity densities in the evanescent wave:

$$w = \gamma I n^2 \omega, \quad \mathbf{p} = \frac{w}{\omega n^2}\left(\frac{k^2}{k_z}\hat{\mathbf{z}} + \sigma\frac{\kappa k}{k_z}\hat{\mathbf{y}}\right), \tag{2.12}$$

$$\mathbf{p}^O = \frac{w}{\omega n^2}\boxed{k_z \hat{\mathbf{z}}}, \quad \mathbf{p}^S = \frac{w}{\omega n^2}\left(-\frac{\kappa^2}{k_z}\hat{\mathbf{z}} + \boxed{\sigma\frac{\kappa k}{k_z}\hat{\mathbf{y}}}\right), \tag{2.13}$$

$$\mathbf{s} = \frac{w}{\omega n^2}\left(\sigma\frac{k}{k_z}\hat{\mathbf{z}} + \boxed{\frac{\kappa}{k_z}\hat{\mathbf{y}}}\right), \quad h = \frac{w}{\omega n}\sigma, \tag{2.14}$$

where $I(x) = |A|^2 \exp(-2\kappa x)$ and $w(x)$ is now the inhomogeneous energy density in the wave. While the helicity and energy densities in Eqs. (2.12) and (2.14) have the same form as for the plane wave in Eqs. (2.8) and (2.10), the momentum and spin densities in Eqs. (2.12)–(2.14) reveal a number of extraordinary features discussed in the main text. As we mentioned, it is the orbital momentum $\mathbf{p}^O$ that represents the observable momentum density in optical fields. In evanescent waves, it corresponds to the '*superluminal*' propagation $k_z > \omega/c$ [shown inside the blue box in Eqs. (2.13)], which can be detected via the anomalously large momentum transfer in both resonant [13] and non-resonant [11,29] light–matter interactions. The *spin* of evanescent waves acquires a *transverse polarization-independent component* $s_y \propto (\kappa/k_z)w$ [shown inside the red box in Eqs. (2.14)], which was predicted recently for surface plasmon-polaritons [30], and which arises from the rotation of the field vectors within the propagation $(x,z)$ plane, see Supplementary Figure 2. Finally, evanescent waves possess *non-zero Belinfante's spin-momentum* density which originates from the presence of spin and inhomogeneous intensity $w(x)$ (see Fig. 2 in the main text). The transverse spin-momentum component determines the *helicity-dependent transverse Poynting momentum* $p_y = p_y^S \propto \sigma(\kappa k/k_z)w$ [shown inside the orange box in Eqs. (2.13)], which was first noticed by Fedorov in 1955 [31]. However, in contrast to the Fedorov's and Imbert's conclusions, this 'virtual' momentum does not lead to energy transport and the standard radiation pressure. Nonetheless, as we show in this work, it can be detected via higher-order light–matter interactions owing to the absence of the transverse orbital momentum.

As we mentioned above, the interaction with real (usually, non-magnetic) particles is highly dual-asymmetric and mostly sensitive to the electric parts of the corresponding field characteristics. Electric and magnetic contributions to the energy, momentum, and spin densities are approximately equal in paraxial propagating fields [2]. However, this is not so for non-paraxial and evanescent fields. Therefore, we also determine separately the electric and magnetic parts (1.17) of the quantities (2.12)–(2.14). This results in

$$w_e = \frac{1}{2}\left(1 + \tau\frac{\kappa^2}{k_z^2}\right)w, \quad w_m = \frac{1}{2}\left(1 - \tau\frac{\kappa^2}{k_z^2}\right)w, \tag{2.15}$$

$$\mathbf{p}_e^O = \frac{w}{2\omega n^2}\boxed{\left(1 + \tau\frac{\kappa^2}{k_z^2}\right)k_z\hat{\mathbf{z}}}, \quad \mathbf{p}_m^O = \frac{w}{2\omega n^2}\left(1 - \tau\frac{\kappa^2}{k_z^2}\right)k_z\hat{\mathbf{z}}, \tag{2.16}$$

$$\mathbf{p}_e^S = \frac{w}{2\omega n^2}\left(-(1+\tau)\frac{\kappa^2}{k_z}\hat{\mathbf{z}} + \boxed{\sigma\frac{\kappa k}{k_z}\hat{\mathbf{y}}}\right), \quad \mathbf{p}_m^S = \frac{w}{2\omega n^2}\left(-(1-\tau)\frac{\kappa^2}{k_z}\hat{\mathbf{z}} + \boxed{\sigma\frac{\kappa k}{k_z}\hat{\mathbf{y}}}\right), \quad (2.17)$$

$$\mathbf{s}_e = \frac{w}{2\omega n^2}\left(\sigma\frac{k}{k_z}\hat{\mathbf{z}} + \boxed{(1+\tau)\frac{\kappa}{k_z}\hat{\mathbf{y}}} + \boxed{\chi\frac{\kappa k}{k_z^2}\hat{\mathbf{x}}}\right), \quad \mathbf{s}_m = \frac{w}{2\omega n^2}\left(\sigma\frac{k}{k_z}\hat{\mathbf{z}} + (1-\tau)\frac{\kappa}{k_z}\hat{\mathbf{y}} - \boxed{\chi\frac{\kappa k}{k_z^2}\hat{\mathbf{x}}}\right). \quad (2.18)$$

These equations reveal several remarkable features. First, the helicity $\sigma$-dependent terms in quantities (2.12)–(2.14) are equally divided into their electric and magnetic parts. Second, the helicity-independent terms of Eqs. (2.12)–(2.14) are asymmetrically divided into electric and magnetic parts depending on the first Stokes parameter $\tau$, Eq. (2.2). This reflects the difference between the electric and magnetic properties of the TM and TE evanescent modes. Finally, the electric and magnetic parts of the spin AM density (2.18) unveil new *vertical* terms $s_{ex} = -s_{mx} \propto \chi(\kappa k/k_z^w)w$ [shown inside the green boxes in Eqs. (2.18)], which are proportional to the degree of the diagonal polarization $\chi$ (the second Stokes parameter). These terms originate from the rotation of the diagonally-polarized electric and magnetic field vectors in the $(y,z)$ plane, see Supplementary Figure 2. Since the electric and magnetic vectors rotate in opposite senses in this plane, the electric and magnetic contributions cancel each other in the total spin density (2.14). However, interaction with an *electric*-dipole particle will reveal the non-zero electric part of this vertical $\chi$-dependent spin density via the vertical radiation torque (see Supplementary Note 3 below).

## Supplementary Note 3. Mechanical action of the fields on small particles

**Analytical calculations for dipole interactions.** A straightforward way to measure the local dynamical characteristic of an optical field (momentum, spin, etc.) is to measure the mechanical action of the field on small probe particles. Therefore, we examine optical forces and torques that appear upon interaction with a small spherical particle. Analytical results can be obtained in the Rayleigh dipole-interaction approximation, when the particle radius $a$ is much smaller than the wavelength: $ka \ll 1$.

A neutral particle in a monochromatic field can be characterized by the complex electric and magnetic dipole moments, $\mathbf{d}_e$ and $\mathbf{d}_m$, induced by the field:

$$\mathbf{d}_e = \alpha_e \mathbf{E}, \quad \mathbf{d}_m = \alpha_m \mathbf{H}, \tag{3.1}$$

where $\alpha_e$ and $\alpha_m$ are the complex electric and magnetic polarizabilities. Using the real dipole moments of the particle, $\mathfrak{d}_e(\mathbf{r},t) = \text{Re}\left[\mathbf{d}_e(\mathbf{r})e^{-i\omega t}\right]$ and $\mathfrak{d}_m(\mathbf{r},t) = \text{Re}\left[\mathbf{d}_m(\mathbf{r})e^{-i\omega t}\right]$ (where $\mathbf{r}$ is the particle's position), the time-averaged optical force $\mathbf{F}$ and torque $\mathbf{T}$ are given by [1,32,33]

$$\mathbf{F} = \left\langle (\mathfrak{d}_e \cdot \nabla)\mathcal{E} + (\mathfrak{d}_m \cdot \nabla)\mathcal{H} + \frac{1}{c}\dot{\mathfrak{d}}_e \times \mathcal{B} - \frac{1}{c}\dot{\mathfrak{d}}_m \times \mathcal{D} \right\rangle, \tag{3.2}$$

$$\mathbf{T} = \left\langle \mathfrak{d}_e \times \mathcal{E} + \mathfrak{d}_m \times \mathcal{H} \right\rangle, \tag{3.3}$$

where $\mathcal{B} = \mu \mathcal{H}$, $\mathcal{D} = \varepsilon \mathcal{E}$, and $\langle ... \rangle$ stands for time averaging. Using Maxwell equations (1.1), we derive expression for the optical force and torque in terms of complex fields and dipole moments:

$$\mathbf{F} = \frac{1}{2}\text{Re}\left[\mathbf{d}_e^* \cdot (\nabla)\mathbf{E} + \mathbf{d}_m^* \cdot (\nabla)\mathbf{H}\right], \tag{3.4}$$

$$\mathbf{T} = \frac{1}{2}\text{Re}\left[\mathbf{d}_e^* \times \mathbf{E} + \mathbf{d}_m^* \times \mathbf{H}\right]. \tag{3.5}$$

Substituting Eq. (3.1) into Eqs. (3.4) and (3.5) and using the expressions (1.2), (1.3), (1.5), and (1.17), we obtain the optical force and torque in terms of the dynamical characteristics of the field [2,15,34–37]:

$$\mathbf{F} = \frac{\gamma^{-1}}{2\omega n^2}\left[\mu \text{Re}(\alpha_e)\nabla w_e + \varepsilon \text{Re}(\alpha_m)\nabla w_m\right] + \gamma^{-1}\left[\mu \text{Im}(\alpha_e)\mathbf{p}_e^O + \varepsilon \text{Im}(\alpha_m)\mathbf{p}_m^O\right], \tag{3.6}$$

$$\mathbf{T} = \gamma^{-1}\left[\mu \text{Im}(\alpha_e)\mathbf{s}_e + \varepsilon \text{Im}(\alpha_m)\mathbf{s}_m\right]. \tag{3.7}$$

The first term in square brackets in Eq. (3.6) describes the *gradient force*, while the second term is the scattering force responsible for *optical pressure*. Thus, the optical pressure is determined by the *orbital momentum density* (1.3) rather than the Poynting vector. For an 'ideal' dual-symmetric particle with $\alpha_e = \alpha_m = \alpha$ in free space, the gradient and scattering radiation forces 'measure' the imaginary and real parts of the *complex* canonical momentum (1.16) of photons [11]:

$$\mathbf{F} = \mathbf{F}_{\text{grad}} + \mathbf{F}_{\text{scat}} = \gamma^{-1}\left[-\text{Re}(\alpha)\text{Im}\,\tilde{\mathbf{p}}^O + \text{Im}(\alpha)\text{Re}\,\tilde{\mathbf{p}}^O\right]. \tag{3.8}$$

Furthermore, the torque (3.7) is proportional to the corresponding electric and magnetic parts of the *spin* density (1.5). The optical pressure and torque are proportional to the imaginary parts of the particle polarizabilities (which are related to the absorption) and to the frequency $\omega$ (in the factor $\gamma^{-1}$). Therefore, this force and torque can be interpreted as the momentum and spin AM transfer rates, from the field to the particle. Moreover, taking into account that the lowest order (in $ka$) term of the polarizability is proportional to $a^3$ (i.e., to the particle's volume), one can conclude that this momentum and spin AM transfer 'measures' meaningful momentum and spin AM densities $\mathbf{p}^O \propto \omega^{-1} \mathbf{F}_{\text{scat}} / a^3$ and $\mathbf{s} \propto \omega^{-1} \mathbf{T} / a^3$. Of course, the particle would 'measure' the proper dual-symmetric field characteristics only in the 'ideal' case of equal electric and magnetic polarizabilities. In practice, they differ significantly (due to the dual asymmetry of matter), and hence the electric and magnetic parts of the field properties are obtained with different efficiencies.

For a spherical particle made of a material with complex permittivity $\varepsilon_p$ and permeability $\mu_p$, the electric and magnetic polarizabilities can be obtained from the Mie scattering coefficients. In the leading orders in $ka$, this results in [38–40]

$$\alpha_e = \frac{\varepsilon}{k^3} \left\{ \frac{\varepsilon_p - \varepsilon}{\varepsilon_p + 2\varepsilon}(ka)^3 + \frac{3}{10} \frac{\varepsilon_p^2 + \varepsilon_p \varepsilon \left[ (\varepsilon_p \mu_p / \varepsilon \mu) - 6 \right] + 4\varepsilon^2}{(\varepsilon_p + 2\varepsilon)^2}(ka)^5 \right\}, \tag{3.9}$$

$$\alpha_m = \frac{\mu}{k^3} \left\{ \frac{\mu_p - \mu}{\mu_p + 2\mu}(ka)^3 + \frac{3}{10} \frac{\mu_p^2 + \mu_p \mu \left[ (\varepsilon_p \mu_p / \varepsilon \mu) - 6 \right] + 4\mu^2}{(\mu_p + 2\mu)^2}(ka)^5 \right\}. \tag{3.10}$$

Usually both the particle and the surrounding medium are *non-magnetic*: $\mu_p = \mu = 1$. This results in the following leading-order polarizabilities (3.9) and (3.10):

$$\alpha_e \simeq \frac{1}{k^3} \frac{\varepsilon(\varepsilon_p - \varepsilon)}{\varepsilon_p + 2\varepsilon}(ka)^3, \quad \alpha_m = \frac{1}{k^3} \frac{(\varepsilon_p - \varepsilon)}{30\varepsilon}(ka)^5. \tag{3.11}$$

In this case, $|\alpha_m| \ll |\alpha_e|$, and in most cases one can consider only electric parts of the forces and torques (3.6) and (3.7), which 'measure' the *electric* parts of the corresponding field characteristics.

Applying the above calculations to the evanescent wave (2.4) and (2.5) with characteristics (2.15)–(2.18), brings about the following results. The longitudinal optical-pressure force will 'measure' the corresponding electric part of the canonical momentum $p^O_{ez}$ [shown inside the blue box in Eq. (2.16)]. The vertical gradient electric-dipole force will indicate the $x$-gradient of the electric energy density $w_e(x)$ [i.e., the 'imaginary' part $\text{Im}\,\tilde{p}^O_{ex}$ of the corresponding complex momentum, see Eqs. (1.16) and (3.8)]. The longitudinal $\sigma$-dependent torque will appear due to the usual helicity-dependent $z$-component of the spin $s_{ez}$. The transverse $\sigma$-independent torque will unveil the helicity-independent $y$-component of the spin $s_{ey}$ [shown inside the red box in Eq. (2.18)]. Finally, the particle will also experience the $\chi$-dependent vertical torque due to the presence of the non-zero $x$-component of the electric spin $s_{ex}$ [shown inside the green box in Eq. (2.18)]. All of these results can be clearly seen in the numerical simulations in Supplementary Figures 4 and 5 and are summarized in Supplementary Table 1.

So far we mostly considered the dipole interactions proportional to the particle volume $a^3$. These interactions are sensitive to the field energy, canonical momentum, and spin densities. At the same time, they do *not* involve *Belinfante's spin momentum* (1.4), which confirms its 'virtual' character. Nonetheless, below we show that the spin momentum, as well as other remarkable quantities, appears in higher-order terms of light–matter interactions. The next-order interaction is the dipole-dipole coupling between the induced electric and magnetic moments. Taking it into account, one can calculate the corresponding force, which is the mixed electric-magnetic force described in [39,40]:

$$\tilde{\mathbf{F}} = \frac{\gamma^{-1}}{3} k^3 \left[ -\mathrm{Re}\left(\alpha_e \alpha_m^*\right) \mathrm{Re}\,\tilde{\mathbf{p}} + \mathrm{Im}\left(\alpha_e \alpha_m^*\right) \mathrm{Im}\,\tilde{\mathbf{p}} \right]. \quad (3.12)$$

Here we introduced the *complex Poynting momentum* $\tilde{\mathbf{p}}$ defined as [1]

$$\tilde{\mathbf{p}} = \frac{g}{2c} \left( \mathbf{E}^* \times \mathbf{H} \right), \quad \mathrm{Re}\,\tilde{\mathbf{p}} = \mathbf{p} = \mathbf{p}^O + \mathbf{p}^S, \quad \mathrm{Im}\,\tilde{\mathbf{p}} = \frac{g}{2c} \mathrm{Im}\left( \mathbf{E}^* \times \mathbf{H} \right). \quad (3.13)$$

Thus, the two terms in the dipole-dipole force (3.12) are proportional to the total Poynting momentum $\mathbf{p}$ (including both the orbital and spin parts) and 'imaginary' Poynting momentum $\mathrm{Im}\,\tilde{\mathbf{p}}$, characterizing an alternating flow of the so-called 'stored energy' [1].

Usually, the spin momentum is accompanied by a non-zero orbital momentum, and the dipole-dipole force (3.12) is negligible compared to the main dipole force (3.6). However, evanescent waves offer a *unique* opportunity to study the *pure* spin transverse momentum $p_y^S = p_y \propto \sigma (\kappa k / k_z) w$ [shown inside the orange box in Eq. (2.13)], without any orbital component. In this case, the transverse dipole force vanishes, and the transverse spin momentum induces *the transverse $\sigma$-dependent dipole-dipole force* (3.12). This offers the *first direct observation* of the fundamental field-theory quantity, introduced in 1939 by Belinfante [7], remarked in optics in 1955 by Fedorov [31], and which was previously considered as 'virtual'. To determine the action of the second term in the force (3.12), we calculate the 'imaginary' Poynting momentum in the evanescent wave (2.4) and (2.5). This yields

$$\mathrm{Im}\,\tilde{\mathbf{p}} = \frac{w}{\omega n^2} \left( -\tau \frac{\kappa k^2}{k_z^2} \hat{\mathbf{x}} - \boxed{\chi \frac{\kappa k}{k_z} \hat{\mathbf{y}}} \right). \quad (3.14)$$

The force (3.12) from the vertical $x$-component $\mathrm{Im}\,\tilde{p}_x$ will be negligible as compared with the dipole gradient force (3.6). At the same time, the transverse component of the 'imaginary' Poynting momentum $\mathrm{Im}\,\tilde{p}_y$ [shown inside the magenta box in Eq. (3.14)] will result in a finite *transverse $\chi$-dependent dipole-dipole force* (3.12) in the diagonally-polarized fields. This force represents a quite intriguing result, namely, a finite optical force at *zero* transverse momentum and intensity gradient: $p_y = p_y^O = p_y^S = 0$ at $\chi = \pm 1$. The presence of the transverse polarization-dependent dipole-dipole forces (3.12) can be clearly seen in the numerical simulations in Supplementary Figures 4 and 5, and these results are summarized in Supplementary Table 1.

For non-magnetic particle with polarizabilities (3.11), the coefficients in the dipole-dipole forces (3.12) take the form:

$$\mathrm{Re}\left(\alpha_e \alpha_m^*\right) \simeq \frac{1}{30 k^6} \left| \frac{\varepsilon_p - \varepsilon}{\varepsilon_p + 2\varepsilon} \right|^2 \left( \mathrm{Re}\,\varepsilon_p + 2 \right) (ka)^8, \quad \mathrm{Im}\left(\alpha_e \alpha_m^*\right) \simeq -\frac{1}{30 k^6} \left| \frac{\varepsilon_p - \varepsilon}{\varepsilon_p + 2\varepsilon} \right|^2 \mathrm{Im}\,\varepsilon_p (ka)^8. \quad (3.15)$$

**Exact numerical calculations for Mie particles in evanescent fields.** Up to now, we described optical forces and torques in the case of small Rayleigh particles, $ka \ll 1$, which allows analytical evaluations and a clear physical interpretation. However, the forces and torques are small in this limit and rapidly grow with $ka$. Therefore, experimental measurements would be more appropriate and feasible when employing Mie particles of moderate size $ka \sim 1$. In this case, the optical forces and torques can be calculated numerically using the exact Mie scattering solutions. Recently, we generalized the Mie scattering solutions for the case of the incident evanescent fields (2.4) and (2.5) [28]. This method is based on the complex-angle rotation (2.3) of the known Mie solutions, and it was approved by comparison with other exact numerical methods. Using the modified Mie procedure of [28], the electric and magnetic fields scattered by a spherical Mie particle in the evanescent wave can be calculated from the evanescent electric field (2.4):

$$\mathbf{E}'(\mathbf{r}) = \hat{M}_E(\mathbf{r})\mathbf{E}, \quad \mathbf{H}'(\mathbf{r}) = \hat{M}_H(\mathbf{r})\mathbf{E}, \tag{3.16}$$

where the matrices $\hat{M}_{E,H}(\mathbf{r})$ include the standard Mie scattering operators and complex-angle rotational operators (2.3). The total electromagnetic field is then given by the sum of the incident and scattered fields:

$$\mathbf{E}^{tot} = \mathbf{E} + \mathbf{E}', \quad \mathbf{H}^{tot} = \mathbf{H} + \mathbf{H}'. \tag{3.17}$$

Once the total field is known, the radiation force can be calculated using the momentum-flux (stress) and the AM-flux tensors, $\hat{\mathcal{T}} = \{\mathcal{T}_{ij}\}$ and $\hat{\mathcal{M}} = \{\mathcal{M}_{ij}\}$:

$$\mathcal{T}_{ij} = \frac{g}{2}\mathrm{Re}\left[\varepsilon E_i^{tot*}E_j^{tot} + \mu H_i^{tot*}H_j^{tot} - \frac{1}{2}\delta_{ij}\left(\varepsilon|\mathbf{E}^{tot}|^2 + \mu|\mathbf{H}^{tot}|^2\right)\right], \quad \mathcal{M}_{ij} = \epsilon_{jkl}x_k\mathcal{T}_{li}, \tag{3.18}$$

where $\epsilon_{ijk}$ is the Levi-Civita symbol, indices take on values $x, y, z$, with $\{x_i\} = (x, y, z)$, and summation over repeating indices is assumed. Integrating the momentum and AM fluxes (3.18) over any surface $\Sigma$ enclosing the particle (e.g., a sphere $\Sigma = \{r = R\}$, $R > a$), we obtain the optical force and torque on the particle:

$$\mathbf{F} = \oint_\Sigma \hat{\mathcal{T}}\mathbf{n}\,d\Sigma, \quad \mathbf{T} = \oint_\Sigma \hat{\mathcal{M}}\mathbf{n}\,d\Sigma, \tag{3.19}$$

where $d\Sigma = R^2 \sin\theta\, d\theta\, d\phi$ is the elementary surface area in spherical coordinates $r\cos\theta = z$, $r\sin\theta\cos\phi = x$, $r\sin\theta\sin\phi = y$, and $\mathbf{n} = (\sin\theta\cos\phi, \sin\theta\sin\phi, \cos\theta)^T$ is the unit vector of the outer normal of the surface $\Sigma$. Note that, as in most other works, in this method we do not account for multiple reflections from the surface limiting the evanescent field. More accurate treatments show that the influence of these reflections is negligible in a wide range of parameters: e.g., in calculations of the force components parallel to the surface, and for particle sizes of the order of the wavelength and not exhibiting resonances [28,29,41–44].

Now, we perform numerical simulations based on the above calculation scheme. For this modelling, we choose the setup and parameters typical for many experiments on evanescent-wave manipulation of Mie particles [45–49]. Namely, we consider the evanescent wave generated via total internal reflection at the interface $x = 0$ between glass (usually, heavy flint glass or sapphire) and water, Supplementary Figure 3. A gold spherical particle is placed in water on the surface of glass, so that its centre is located at $x_p = a$. The generation of the evanescent wave in the total reflection is described by Eqs. (2.6) and (2.7), and all materials are non-magnetic, $\mu_1 = \mu = \mu_p = 1$. The other parameters are:

$$\varepsilon_1 = 3.06, \quad \varepsilon = 1.77, \quad \varepsilon_p = -12.2 + 3i, \quad \theta = 51° \ (\theta_c = 49.5°), \quad \kappa/k = 0.21, \tag{3.20}$$

and the wavelength in vacuum is assumed to be $\lambda_0 = 2\pi c/\omega = 650\,\mathrm{nm}$. Using these parameters, we calculate all components of the radiation force and torque on the particle for six basic polarizations of the evanescent field: $\tau = \pm 1$, $\chi = \pm 1$, and $\sigma = \pm 1$. The results, as functions of the particle size $ka$, are presented in Supplementary Figure 4, where the forces and torques are normalized by the following quantities:

$$F_0 = \frac{a^2}{4\pi}|A_1|^2, \quad T_0 = \frac{F_0}{k}, \tag{3.21}$$

Here the normalization factor involves the square of the particle size to improve the visibility of the data for both small- and moderate-size particle (as it was also used in [28,41]), and we recall that $A_1$ is the amplitude of the incident plane wave in the glass.

Supplementary Figure 4 shows the presence of all forces and torques (3.6), (3.7), (3.12), which quantify the four distinct momenta and three distinct spins in characteristics (2.12)–(2.18) and (3.14) of the evanescent field. This correspondence is summarized in Supplementary Table 1. Most importantly, the $\sigma$-independent torque $T_y$ indicates the transverse helicity-independent spin in the evanescent wave [shown inside the red box in Eqs. (2.14) and (2.18)]. Next, the vertical $\chi$-dependent torque $T_x$ reveals the presence of the vertical electric spin in the diagonally-polarized evanescent waves [shown inside the green box in Eq. (2.18)]. Finally, the $\sigma$- and $\chi$-dependent transverse forces $F_y$ unveil the presence of the helicity-dependent transverse spin momentum [shown inside the orange boxes in Eqs. (2.13) and (2.17)] and 'imaginary' transverse Poynting vector [shown inside the magenta box in Eqs. (3.14)]. Note that these transverse forces are one order of magnitude weaker than typical radiation forces, $F_z$ and $F_x$, and they vanish in the Rayleigh-particle limit $ka \ll 1$. Although the analytical expressions for the forces and torques, (3.6), (3.7), (3.12), are derived in the $ka \ll 1$ approximation, the exact forces and torques in Supplementary Figure 4 show the same polarization dependence and qualitative picture for larger particles with $ka \sim 1$.

In Supplementary Figure 5 we compare the exact numerical calculations of Supplementary Figure 4 with approximate analytical expression for forces and torques on a Rayleigh particle with $ka \ll 1$, Eqs. (3.6)–(3.15). One can see that the dipole and dipole-dipole weak-coupling approximations describe the leading orders of the forces and torques in the Rayleigh $ka \ll 1$ limit, but the exact forces and torques usually become larger in the strong-coupling Mie region with $ka \sim 1$.

Thus, we have shown that evanescent electromagnetic waves can carry *four* distinct momenta and *three* distinct spin angular momenta. This is in sharp contrast with the single momentum and single spin for a propagating plane wave (photons). Each of these momenta and spins has a clear physical meaning and result in a corresponding *directly-observable* force or torque on a probe Mie particle, as shown in Supplementary Figure 4. The field characteristics are given in Eqs. (2.12)–(2.18) and (3.14), whereas the forces and torques are described by Eqs. (3.6), (3.7), (3.11), and (3.12) in the $ka \ll 1$ approximation. These results are summarized in Supplementary Table 1, which shows excellent agreement with the exact numerical simulations in Supplementary Figure 4.

# Supplementary References